\pdfoutput=1

\documentclass[12pt,manuscript]{aastex}
\usepackage{color}
\usepackage{amsmath}
\usepackage{graphicx}
\usepackage{txfonts}
\usepackage{CJK}

\newcommand{\ym}{}
\newcommand{\ymone}{}

\slugcomment{}

\shorttitle{CSTAR Binaries}
\shortauthors{Yang et al.}

\begin{document}

\begin{CJK*}{UTF8}{gbsn}

\title{ECLIPSING BINARIES FROM THE CSTAR PROJECT \\ AT DOME A, ANTARCTICA}

\author{Ming Yang {\ym (杨明)}\altaffilmark{1},
Hui Zhang\altaffilmark{1,2},
Songhu Wang\altaffilmark{1},
Ji-Lin Zhou\altaffilmark{1,2},
Xu Zhou\altaffilmark{3,4},
{\ym Lingzhi Wang\altaffilmark{3}},\\
Lifan Wang\altaffilmark{5},
R. A. Wittenmyer\altaffilmark{6},
Hui-Gen Liu\altaffilmark{1},
Zeyang Meng\altaffilmark{1},
M. C. B. Ashley\altaffilmark{6},\\
J. W. V. Storey\altaffilmark{6},
D. Bayliss\altaffilmark{7},
Chris Tinney\altaffilmark{6},
Ying Wang\altaffilmark{1},
Donghong Wu\altaffilmark{1},
Ensi Liang\altaffilmark{1},\\
Zhouyi Yu\altaffilmark{1},
Zhou Fan\altaffilmark{3,4},
Long-Long Feng\altaffilmark{5},
Xuefei Gong\altaffilmark{8},
J. S. Lawrence\altaffilmark{6,9},
Qiang Liu\altaffilmark{3},\\
D. M. Luong-Van\altaffilmark{6},
Jun Ma\altaffilmark{3,4},
Zhenyu Wu\altaffilmark{3,4},
Jun Yan\altaffilmark{3},
Huigen Yang\altaffilmark{10},
Ji Yang\altaffilmark{5},\\
Xiangyan Yuan\altaffilmark{8},
Tianmeng Zhang\altaffilmark{3,4},
Zhenxi Zhu\altaffilmark{5}, Hu Zou\altaffilmark{3,4}
}

\altaffiltext{1}{School of Astronomy and Space Science, Key Lab of Modern Astronomy and Astrophysics,  Nanjing University, Nanjing 210046, China}
\altaffiltext{2}{Collaborative Innovation Center of Modern Astronomy and Space Exploration,  Nanjing 210046, China}
\altaffiltext{3}{National Astronomical Observatories, Chinese Academy of Sciences, Beijing 100012, China}
\altaffiltext{4}{Key Laboratory of Optical Astronomy, National Astronomical Observatories, Chinese Academy of Sciences, Beijing 100012, China}
\altaffiltext{5}{Purple Mountain Observatory, Chinese Academy of Sciences, Nanjing 210008, China}
\altaffiltext{6}{School of Physics, University of New South Wales, NSW 2052, Australia}
\altaffiltext{7}{Research School of Astronomy and Astrophysics, Australian National University, Canberra, ACT 2611, Australia}
\altaffiltext{8}{Nanjing Institute of Astronomical Optics and Technology, Nanjing 210042, China}
\altaffiltext{9}{Australian Astronomical Observatory, NSW 1710, Australia}
\altaffiltext{10}{Polar Research Institute of China, Pudong, Shanghai 200136, China}

\email{huizhang@nju.edu.cn; zhoujl@nju.edu.cn}

\begin{abstract}
	The Chinese Small Telescope ARray (CSTAR) has observed an area around the Celestial South Pole at Dome A since 2008. About $20,000$ light curves in the i band were obtained lasting from March to July, 2008. The photometric precision achieves about 4 mmag at i = 7.5 and 20 mmag at i = 12  within a 30 s exposure time. These light curves are analyzed using Lomb--Scargle, Phase Dispersion Minimization, and Box Least Squares methods to search for periodic signals. False positives may appear as a variable signature caused by contaminating stars and the observation mode of CSTAR. Therefore the period and position of each variable candidate are checked to eliminate false positives. Eclipsing binaries are removed by {\ymone visual inspection, frequency spectrum analysis and locally linear embedding technique.} We identify $53$ eclipsing binaries in the field of view of CSTAR, containing $24$ detached binaries, $8$ semi-detached binaries, $18$ contact binaries, {\ymone and 3 ellipsoidal variables}. To derive the parameters of these binaries, we use the Eclipsing Binaries via Artificial Intelligence (EBAI) method. The primary and the secondary eclipse timing variations (ETVs) for semi-detached and contact systems are analyzed. Correlated primary and secondary ETVs confirmed by false alarm tests may indicate an unseen perturbing companion. Through ETV analysis, we identify two triple systems (CSTAR J084612.64-883342.9 and CSTAR J220502.55-895206.7). The orbital parameters of the third body in CSTAR J220502.55-895206.7 are derived using a simple dynamical model. 
\end{abstract}


\section{INTRODUCTION}
Binaries have made great contributions to stellar fundamental parameters and evolutionary models. Eclipsing binaries are systems whose component stars eclipse mutually along the line of sight to the observer. Light curves of eclipsing binaries contain information on orbital inclination, eccentricity, brightness ratio, relative stellar sizes, etc. Mass and radius can be determined to high accuracy by combining radial velocity and multi-band photometry \citep{and91}. With accurate fundamental parameters, stellar structure and evolution theories can be tested \citep{pol97,gui00,whi01,tor02}. The analysis of eclipsing binaries can help us to  understand many astrophysical problems, e.g. the O'connell effect \citep{oco51,mil68,dav84}, which refers to the different maxima in brightness of some binary light curves; and the Algol Paradox \citep[reviewed by][]{pus05}, which refers to the phenomena that binaries seem to evolve in discord with the established theories of stellar evolution. Therefore eclipsing binaries contribute to various fields of astronomy.

The solution of an eclipsing binary light curve is a mature field. \citet{kal99} reviewed some important physical models and codes. The most widely used one is the WD code \citep{wil71}. It is also the engine of PHOEBE \citep[PHysics Of Eclipsing BinariEs,][]{prs05} package. PHOEBE is accurate but time-consuming when data volumes grow and the number of light curves increases. Compared to PHOEBE, EBAI \citep[Eclipsing Binaries via Artificial Intelligence,][]{prs08} method is time-saving. It will learn from the modeled eclipsing binary light curves generated by PHOEBE, then recognize parameters of unknown eclipsing binaries. EBAI is appropriate for wide-field photometric surveys especially when the data volumes are very large. To calculate eclipsing binary parameters automatically and efficiently, we choose the EBAI pipeline.

More and more projects provide astronomers chances to find eclipsing binaries, e.g. the OGLE \citep[Optical Gravitational Lensing Experiment;][]{uda97}, MACHO \citep[Massive Astrophysical Compact Halo Objects;][]{alc97}, ASAS \citep[All Sky Automated Survey;][]{poj02} and \textit{Kepler} \citep{bor04}. \textit{Kepler} is one of the most successful space telescopes, which was launched in 2009 and has already found $2165$ eclipsing binaries \citep{prs11,sla11,mat12,conroy13}. Circumbinary planets have also been confirmed in several {\it Kepler} systems \citep{doy11,wel12,oro12a,oro12b}. The success of \textit{Kepler} should be attributed to the steady space conditions and its continuous observation, which is unmatched by ground-based surveys.

In the past several years, the Antarctic plateau attracts the attention of many astronomers. It is extremely cold and dry, and has continuous polar nights. Dome A is located at longitude $77^{\circ}06^{'}57^{''}$E, latitude $80^{\circ}25^{'}08^{''}$S, $4093$m above the sea level. All the results of site testing indicate that it has great potential for astronomical observations \citep{law08,law09,yan09,sau09,zou10}. In 2008, the Chinese Small Telescope ARray (CSTAR) was installed in Dome A and a lot of scientific data have been returned since then. Previous works based on the CSTAR data have corrected some systematic effects and found many variable stars \citep{wlz11,wlz13,wang12,wang14a,meng13}. In this paper, we present our work on identifying eclipsing binaries from the CSTAR data.

This paper is arranged as follows. Section 2 describes instruments, the strategy of observations and data preparation. Section 3 shows the methods of searching eclipsing binaries and gives the eclipsing binary catalog. Section 4 computes parameters for different types of eclipsing binaries. Section 5 analyzes the ETVs of semi-detached and contact binaries. Section 6 presents parameter distributions and discusses several interesting systems. Section 7 concludes our work.

\section{INSTRUMENTS AND OBSERVATIONS}
CSTAR is the first Antarctic telescope array designed and constructed by China. It contains four Schmidt-Cassegrain telescopes. Each CSTAR telescope has a pupil entrance aperture of $14.5\ \rm{cm}$ with a focal ratio of $\rm{f}/1.2$. The small aperture allows CSTAR to cover a large field of view of $4^{\circ}.5 \times 4^{\circ}.5$. Each focal plane has a $1\ \rm{k} \times 1\ \rm{k}$ frame-transfer CCD with a pixel size of $13\ \rm{\mu m}$, giving a plate scale of $15\arcsec/\rm{pix}$. Three of the CSTAR telescopes have fixed filters: $g,\ r$, and $i$, similar to those used by the Sloan Digital Sky Survey (SDSS). Their effective wavelengths are $470\ \rm{nm},\ 630\ \rm{nm}$, and $780\ \rm{nm}$ \citep[see][table~1]{zho10b}. The fourth telescope has no filter. All the telescopes are fixed pointing at the direction near the Celestial South Pole. Therefore, stars travel circularly in the FOV.

After tested at Xinglong Observatory in September 2007, CSTAR was shipped to Antarctica and commissioned at Dome A in January 2008. CSTAR has no mechanical shutter to minimize the risk. The exposure time was $20$ s before 2008 April 4 and $30$ s thereafter. {\ym No useful data in the  g, r, and open bands were obtained because of intermittent problems with the CSTAR computers and hard disks \citep{yan09}.} Fortunately one telescope with $i$-band filter worked well from 2008-03-20 to 2008-07-29. In polar nights observations were continuous for twenty-four hours. When the solar elevation angle gradually increased, observation error increased and decreased diurnally. Technical problems resulted in two gaps in the observations, of 10 days and 15 days. Finally more than $287,800$ images were taken with a total integration time of $1,615$ hours in 2008 \citep{zho10a}.

Preliminary image processing and photometry include bias subtraction, flat-field, and fringe correction \citep{zho10a,zho10b}. Dark current is negligible under low temperature conditions. Due to the continuous, shutterless observation mode, there are no real-time bias or daily flat field frames. They were created by using the images obtained during the four test-observation nights in Xinglong Observatory from September $3^{\rm{rd}}$ to $7^{\rm{th}}$, 2007. The variations of flat-field are more complicated due to different observation sky areas and the lower temperature of Antarctica. Fortunately, it can be corrected by using all of the circular traces of the stars\citep[see][figure~8]{zho10b}.

The USNO-B1.0 catalog \citep{mon03} contains well-calibrated magnitudes of the point sources in the observed field of CSTAR. \citet{mon03} have derived a transformation from USNO-B1.0 magnitudes to SDSS magnitudes. Because filters of CSTAR are similar to those of SDSS, USNO-B1.0 catalog can be used to determine the photometric calibration directly. Time calibration was taken by using the position of each star as a clock. Corrected Julian date (JD) at the mid-exposure point of every image was presented in each catalog. The accuracy can reach several seconds \citep{zho10a}.

\citet{wang12} correct the inhomogeneous effect of cloud on CSTAR photometry, including the high cirrus and the fog near the ground surface. \citet{meng13} correct the ghost images, which is caused by the Schmidt-Cassegrain optical structure. {\ym As CSTAR was fixed pointing to the Celestial South Pole, daily stellar movements in CCD will cause diurnal variation for each light curve due to the CCD unevenness. It has been removed by \citet{wang14a}. After these corrections, the photometric precision can reach about 4 mmag at i=7.5 and 20 mmag at i=12 \citep{wang14b}. About 20,000 sources down to 16 mag were detected. The revised CSTAR catalog and data of 2008 are available at http://explore.china-vo.org. The following work is based on the detrended light curves after these corrections}.

\section{ECLIPSING BINARY CATALOG}
{\ym Eclipsing binaries are not easy to be distinguished from other kinds of variables. Therefore it is necessary to automatically search variables at first and then manually select out eclipsing binaries from the variables. However, false positives caused by contaminating stars and the observation mode of CSTA may appear variable signature. They should be removed from the variable candidates before confirming eclipsing binaries. In this section, we describe the design of the eclipsing binary catalog.}

\subsection{Searching Periodic Signals}
In the first step, periodic signals are extracted from each light curve. A bin size of five minutes has been adopted to filter out extremely high frequency noises because CSTAR has a very short cadence. Periodic signals are recognized using {\ym three} methods: Lomb-Scargle \citep{lom76,sca82}, Phase Dispersion Minimization \citep[PDM;][]{ste78,sch89}, {\ym and Box Least Squares \citep[BLS;][]{kov02}}. We set the range of period scan from 0.1 to 30 days {\ym for PDM and BLS method. While for Lomb-Scargle method the lower limit is increased to 1.05 day. For each light curve, we calculate its Lomb-Scargle signal-to-noise ratio R, PDM statistic $\Theta$ \citep[see][Equation (3)]{ste78}, and Signal Detection Efficiency \textit{SDE} \citep[see][Equation (6)]{kov02}. A false alarm probability of $10^{-4}$ is assigned to Lomb-Scargle method to get the power threshold $\sigma_{\rm LS}$. We set 
$$R=S_{\rm LS}/\sigma_{\rm LS}$$
where $S_{\rm LS}$ is the power of the highest peak in the Lomb-Scargle periodogram. Higher R value indicates more significant periodic signal. $\Theta$ is between 0 and 1. Lower $\Theta$ value indicates more significant periodic signal. \textit{SDE} can reflect the effective signal-to-noise ratio of eclipses. It has been discussed detailedly by \citet{kov02}. We choose the criteria
$$R_{\rm c}=10,~\Theta_{\rm c}=0.85,~{\it SDE}_{\rm c}=6$$
where the subscript ``c'' represents ``criteria''. Lomb-Scargle, PDM, and BLS methods contribute 225, 107, and 244 variable candidates, respectively.   93 of them are detected by two methods and 27 are detected by all the three methods. Therefore there are totally 429 variable candidates when any one of the condition is met.}

\subsection{Rejecting False Positives}
False positives come from four sources. The first one is the diurnal variation. Though very weak after detrending, it still should be taken into consideration. Secondly, stars at the edges of the FOV have regular gaps because stars move in and out of the FOV repeatedly every day. It may cause additional periodic variations. Thirdly, an exposure time of $20$ seconds or $30$ seconds is too long for a bright variable star, which can pollute its neighborhood. A typical phenomenon is the fact that they appear to have the same variations. Therefore when two or more targets show nearly the same periods, it is necessary to check if they are very close and if their light curves vary simultaneously. Additionally, a large plate scale of $15\arcsec / \rm{pix}$ may also produce false eclipses {\ym induced by} a nearby non-variable star. Other sources of contamination include solar brightness or moonlight at some epoches and occasional aurora, all of which can brighten the sky background. Images with high background have been discarded in the data reduction process. Therefore such contaminations have little effects.

All variable stars are checked in three ways. Their positions, periods and variable trends are compared. False positives are confirmed if any one of the following conditions is met:
\begin{itemize}
\item Distance less than ten pixels to a saturated star.
\item Period equals one Sidereal Day or diurnal harmonics.
\item Variable trend resemble that of its neighborhood.
\end{itemize}

\subsection{Classification of Eclipsing Binaries}

\begin{figure}
\epsscale{0.8}
\plotone{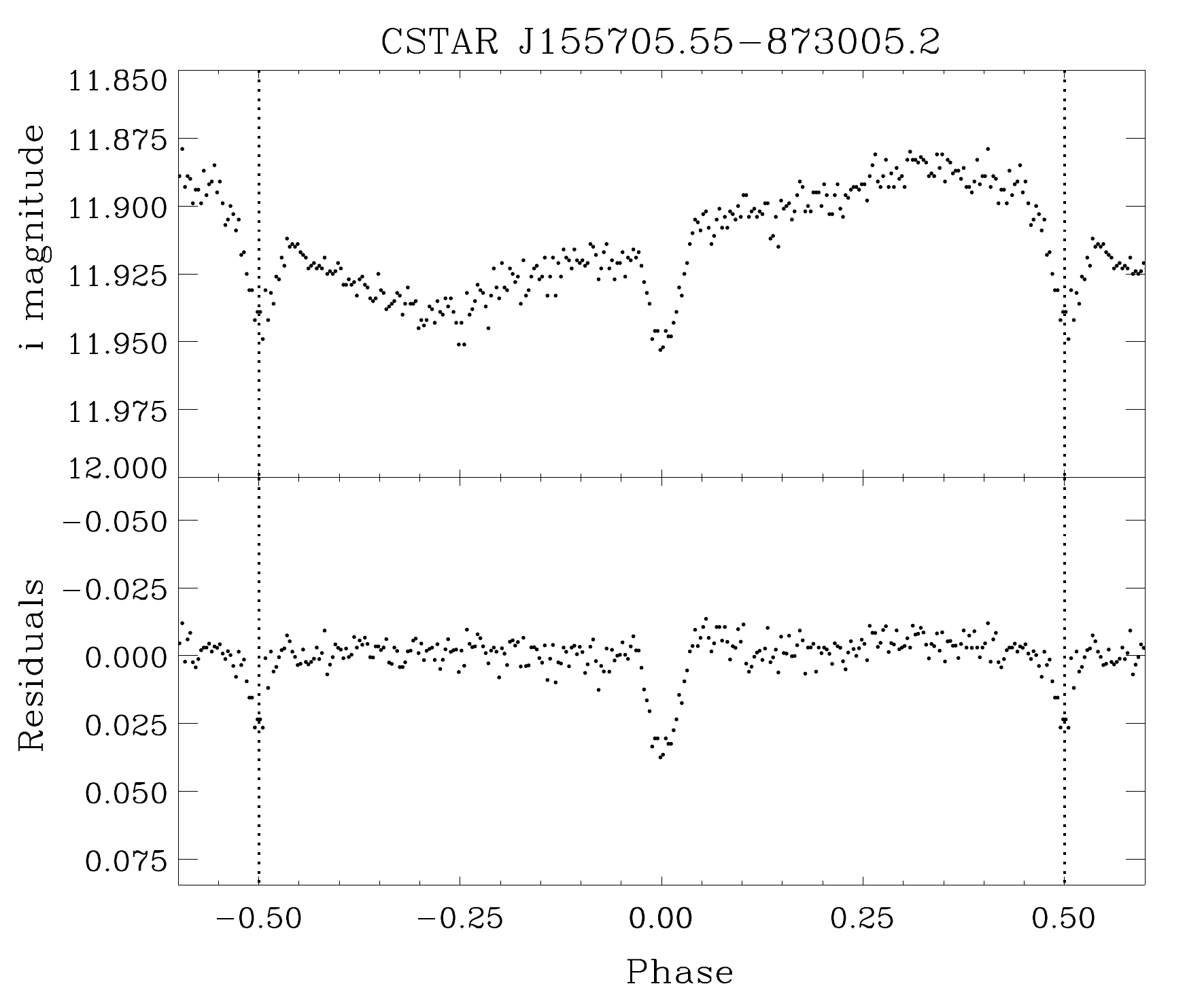}
\caption{Detached RS Canum Venaticorum variable in the CSTAR FOV. The top panel is the folded and binned light curve. The bottom panel is the residuals after removing the strongest periodic signal. The period of the system is 3.111666 $\pm$ 0.000207 days.
\label{rs}}
\end{figure}

All remaining light curves after culling false positives  are manually inspected to exclude variables with similar shapes as eclipsing binaries, e.g. $\gamma$ Doradus, $\delta$ Scuti, RR Lyrae, etc. To check if there exist eclipsing binaries in other types of variable stars, we subtract the strongest period of each variable star and analyze the residuals using the same method described above. A detached RS Canum Venaticorum variable is found as shown in Figure \ref{rs}. 

The visual inspection of variables is carried out by two groups of the authors (M. Yang et al. and S. Wang et al.) independently. Eclipsing binaries are selected out and classified into four types according to their morphologies \citep{pac06,prs11}:
\begin{itemize}
\item[1.] Eclipsing Detached binary (ED) -- neither component fills its Roche Lobe;
\item[2.] Eclipsing Semi-Detached binary (ESD) -- only one component fills its Roche Lobe;
\item[3.] Eclipsing Contact binary (EC) -- both components fill their Roche lobes;
{\ymone \item[4.] Ellipsoidal variable (ELL) -- low-inclination binaries with close ellipsoidal components.}
\end{itemize}

\begin{figure}
\epsscale{0.7}
\plotone{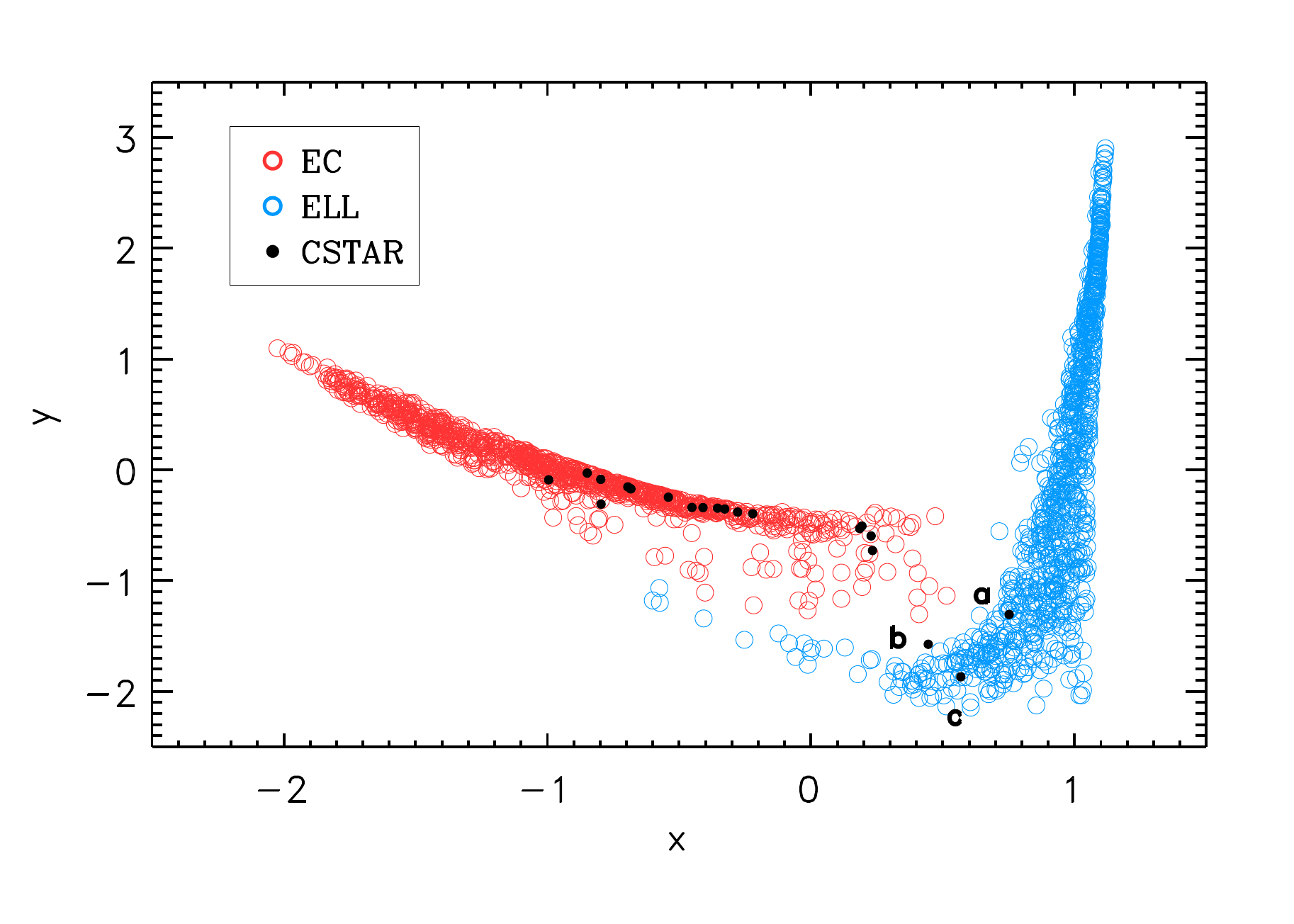}
\caption{ The two-dimensional LLE projection of the EC and ELL light curve space. Red and blue circles are sampled EC and ELL light curves, respectively. Black dots are light curves from CSTAR. Three CSTAR light curves fall into the ELL region. They are (a)CSTAR J100121.80-881330.8 (b)CSTAR J022530.81-871311.9 and (c)CSTAR J191753.08-885111.2. Note that the new coordinates (x,y) of each light curve do not depend on global translations, rotations and scalings.
\label{lle}}
\end{figure}

ED light curves are nearly flat-topped with separate eclipses. ESD light curves are continuously variable with a large difference in depth between the primary and the secondary eclipses. Therefore detached and semi-detached systems are easy to be recognized due to their distinct eclipses. However, for a contact system with indistinct ingress and egress points of the eclipses, visual inspection is not reliable. {\ymone Challenges mainly come from $\delta$ Scuti variables and ELLs. Because both $\delta$ Scuti and ELL light curves exhibit sinusoidal variations. If a contact system appears approximately equal primary and secondary eclipses, it is difficult to distinguish the EC light curve from $\delta$ Scuti and ELL light curves. We take methods of frequency spectrum analysis and Locally Linear Embedding \citep[LLE;][]{RS00} technique to solve the problem.

The frequency spectrums of ECs and $\delta$ Scutis are different. $\delta$ Scuti light curves exhibit variations due to both radial and non-radial pulsations of the star's surface. Therefore the frequency spectrum of a $\delta$ Scuti usually contains more peaks caused by the multi-mode pulsations. Whereas the frequency spectrum of an EC light curve contains only one strong signal and harmonics of the signal. For CSTAR data with a short cadence of $\sim$30 s, the harmonics are usually very weak. We calculate the frequency spectrums of all sinusoidal variables using Lomb-Scargle method from 0.025 day to 0.95 day for further reference.
	
ELL light curves exhibit sinusoidal variations due to the changing emitting area toward the observer. We adopt the LLE method proposed by \citet{mat12} to distinguish ECs and ELLs. LLE is a nonlinear dimensionality reduction technique. This method has been applied to {\it KEPLER} data to classify eclipsing binary light curves and it turns out to be successful.  LLE can remember the local geometry of a higher dimensional data set, and reconstruct a lower dimensional projection with the same local geometry. Therefore light curves with similar features will stay adjacent to each other in the lower dimensional projection. We generate 1000 EC light curves and 1000 ELL light curves using PHOEBE package, respectively. The amplitudes of these light curves are scaled to the [0,1] interval. The light curves are sampled at 100 equidistant phases; in other words, each light curve can be treated as a point in a $D=100$ dimensional space. To preserve the local geometry, every light curve is characterized by a linear combination of its $k=20$ neighbors. We reduce the $D=100$ dimensional space to a $d=2$ dimensional space. The final two-dimensional LLE projection is shown in Figure \ref{lle}. Each point represents a light curve sampled at 100 equidistant phases. Sampled EC light curves (red circles) are clustered in the red region, and sampled ELL light curves (blue circles) in blue region. Three light curves fall into the ELL region. They are listed in Table \ref{ell_pars}. The periods and ephemerides of the ELLs are derived in Section 4.}

\begin{deluxetable}{cccc}
\tabletypesize{\scriptsize}
\tablewidth{0pt}
\tablecaption{Ellipsoidal variables.}
\tablehead{
\colhead{CSTAR ID} & \colhead{mag}&\colhead{HJD}  & \colhead{Period} \\
   \colhead{}& \colhead{}&\colhead{(JD-2454500)}  & \colhead{(days)}
}
\startdata
CSTAR J022530.81-871311.9	&	13.349	($\pm$	0.031	)&	49.964260	($\pm$	0.006070	)&	0.456869	($\pm$	0.000035	)\\
CSTAR J100121.80-881330.8	&	11.855	($\pm$	0.014	)&	49.311958	($\pm$	0.000269	)&	0.652239	($\pm$	0.000002	)\\
CSTAR J191753.08-885111.2	&	11.824	($\pm$	0.014	)&	49.339306	($\pm$	0.000433	)&	0.372034	($\pm$	0.000002	)\\
\enddata
\label{ell_pars}
\begin{center}
{\bf Note.} Columns 1-4 represent CSTAR ID, magnitude, the reference time of primary minimum and period. J represents J2000.0.
\end{center}
\end{deluxetable}

{\ymone Combining visual inspection, frequency spectrum analysis, and LLE technique, eclipsing binaries are selected out and voted by all members. Finally $53$ systems are classified into $24$ EDs ($45\%$), $8$ ESDs ($15\%$), $18$ ECs ($34\%$), and $3$ ELLs ($6\%$).} Compared with other projects, the fractions of different types of eclipsing binaries from the University of New South Wales (UNSW) Extrasolar Planet Search are $43.1\%$ for EDs, and ESDs, and $56.9\%$ for ECs \citep{chr08}; $Kepler$ are $58.2\%$ for EDs, $7\%$ for ESDs, $21.7\%$ for ECs, $6.3\%$ for ELLs, and $6.8\%$ for uncertain types \citep{sla11}. Longer observation time and unprecedented precision of $Kepler$ make it sensitive to find long-period EDs. Our result lies between the two projects.

\section{ECLIPSING BINARY PARAMETERS}

{\ym In this section we give the binary parameters as shown in Tables \ref{bound}-\ref{ec_pars}. Two methods are applied to determine the accurate periods and ephemerides. The first one is the classical O-C method. Epochs of minimum light are given by K-W method \citep{KW56}. Then a linear fit of the epoches is performed to derive the period and ephemeris, as described by \citet{zha14}. For the second method, prior to the linear fit, we adopt a polynominal fit to determine the epoches of the light minima instead of the K-W method. Details are described in section 5.1. These two methods are performed separately and the results with higher precisions are adopted.} 

{\ym With the previous obtained periods and ephemerides, physical} parameters of these eclipsing binaries are computed using the EBAI method. EBAI introduces artificial neural networks to learn the characteristics by training on large data sets. Then the knowledge is applied to recognize physical parameters of new eclipsing binaries. \citet{prs08} have described the principles of the method. Test results of applying it to EDs from CALEB and OGLE database point to significant viability. \citet{prs11} and \citet{sla11} have discussed how to choose principal parameters for ED, ESD, and EC \citep[see][Table~2]{prs11}. We use the parameters they recommended. First, we ran the PHOEBE package to generate modeled eclipsing binary light curves for the training process. PHOEBE is a modeling package for eclipsing binaries based on Wilson-Devinney program. It retains all Wilson-Devinney codes as the lowermost layer, the extension of physical models and technical solutions as the intermediate layer, and the user interface as the topmost layer. Parameters for different types of eclipsing binaries are calculated with different methods.

\subsection{ Parameters of Detached and Semi-detached Binaries }
\begin{figure}
\epsscale{0.7}
\plotone{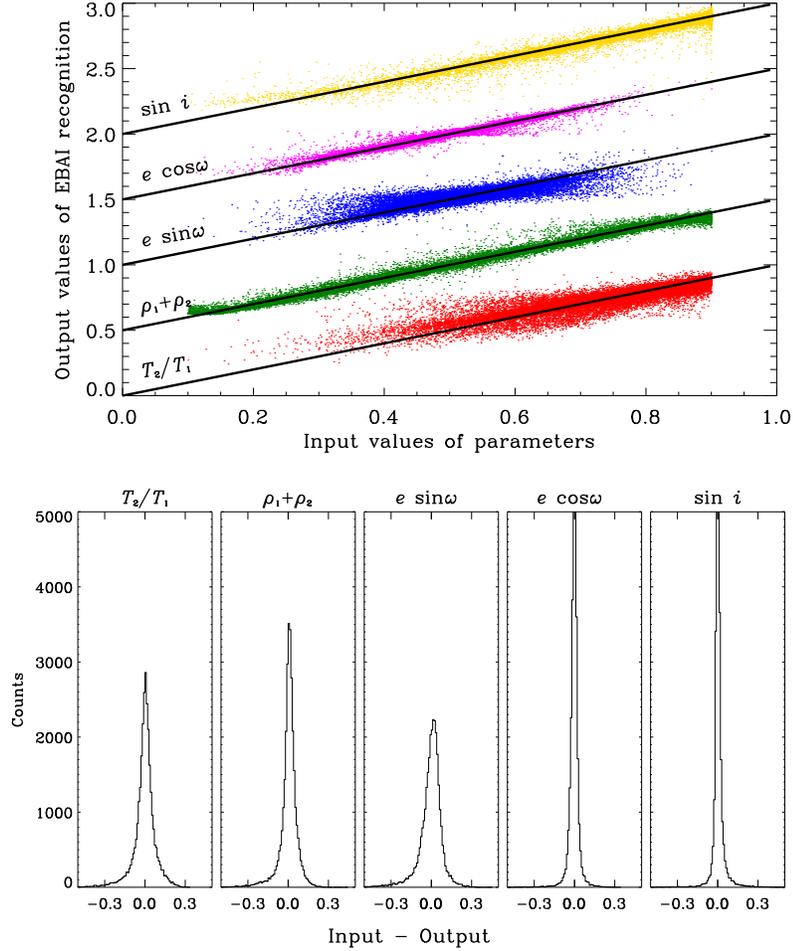}
\caption{ EBAI performance and parameter error distributions of simulated detached 
and semi-detached binaries. For accurate calculation, all the original parameters are linearly scaled to the [0.1,0.9] interval by EBAI. The ranges of the original parameters are shown in Table \ref{bound}. Top panel: Recognition results for 30,000 exemplars. The X axis represents input values of parameters, and the Y axis represents the output values after recognition by EBAI. The parameters are offset by 0.5 in the Y axis for clarity.  Bottom panels: Error distributions of the parameters. The parameter errors are obtained by comparing the input values and the output values in the top panel.
\label{lt_ed}}
\end{figure}

For EDs and ESDs, we choose five principal parameters: the temperature ratio $T_{\rm{2}}\ /\ T_{\rm{1}}$, which determines the eclipse depth ratio; the sum of fractional radii $\rho_{\rm{1}}+\rho_{\rm{2}}$, which determines eclipse width; the eccentricity $e$ and the argument of periastron $\omega$ in orthogonal forms $e\cdot\rm{\sin}\omega$ and $e\cdot\rm{\cos}\omega$, which determine the separation between primary and secondary eclipse; the sine of inclination $\sin i$, which determines the eclipse shape. 
We generate $30,000$ simulated light curves by randomly sampling the five parameters as a training set for EBAI. After $200,000$ training iterations, a correlation matrix which can recognize parameters from eclipsing binary light curves is generated. We test the correlation matrix with $30,000$ unknown light curves. The recognizing results and the parameter error distributions are shown in Figure \ref{lt_ed}. All the parameters have been linearly scaled to the [0.1,0.9] interval by EBAI. The ranges of the parameters are shown in Table \ref{bound}.

\begin{deluxetable}{cccccccccc}
\tabletypesize{\scriptsize}
\tablewidth{0pt}
\tablecaption{ Ranges of parameters.  }
\tablehead{
\multicolumn{5}{c}{ED/ESD} & \colhead{} & \multicolumn{4}{c}{EC} \\
\cline{1-5} \cline{7-10}
\colhead{$T_2/T_1$} & \colhead{$\rho_1+\rho_2$} & \colhead{$e$} & \colhead{$\omega$} & \colhead{$i$} & &
\colhead{$T_2/T_1$} & \colhead{$M_2/M_1$} & \colhead{$Fillout$} &  \colhead{$i$} \\
\colhead{} & \colhead{} & \colhead{} & \colhead{($^\circ)$} & \colhead{$(^\circ)$} & &
\colhead{} & \colhead{} & \colhead{} &  \colhead{$(^\circ)$} 
}
\startdata
[0.1,1.0] & [0.05,0.75] & [0,1] & [0,360] & [60,90] &  & [0.5,1.0] & [0.1,5.0] & [0,1] & [35,90] 
\enddata
\label{bound}
\begin{flushleft}
{\bf Note.} Columns 1-5 are parameter ranges of EDs and ESDs: temperature ratio, the sum of fractional radii, eccentricity, the argument of periastron, and inclination. Columns 6-9 are parameter ranges of ECs: temperature ratio, mass ratio, fillout factor, and inclination.
\end{flushleft}
\end{deluxetable}

\begin{figure}
\epsscale{0.9}
\plotone{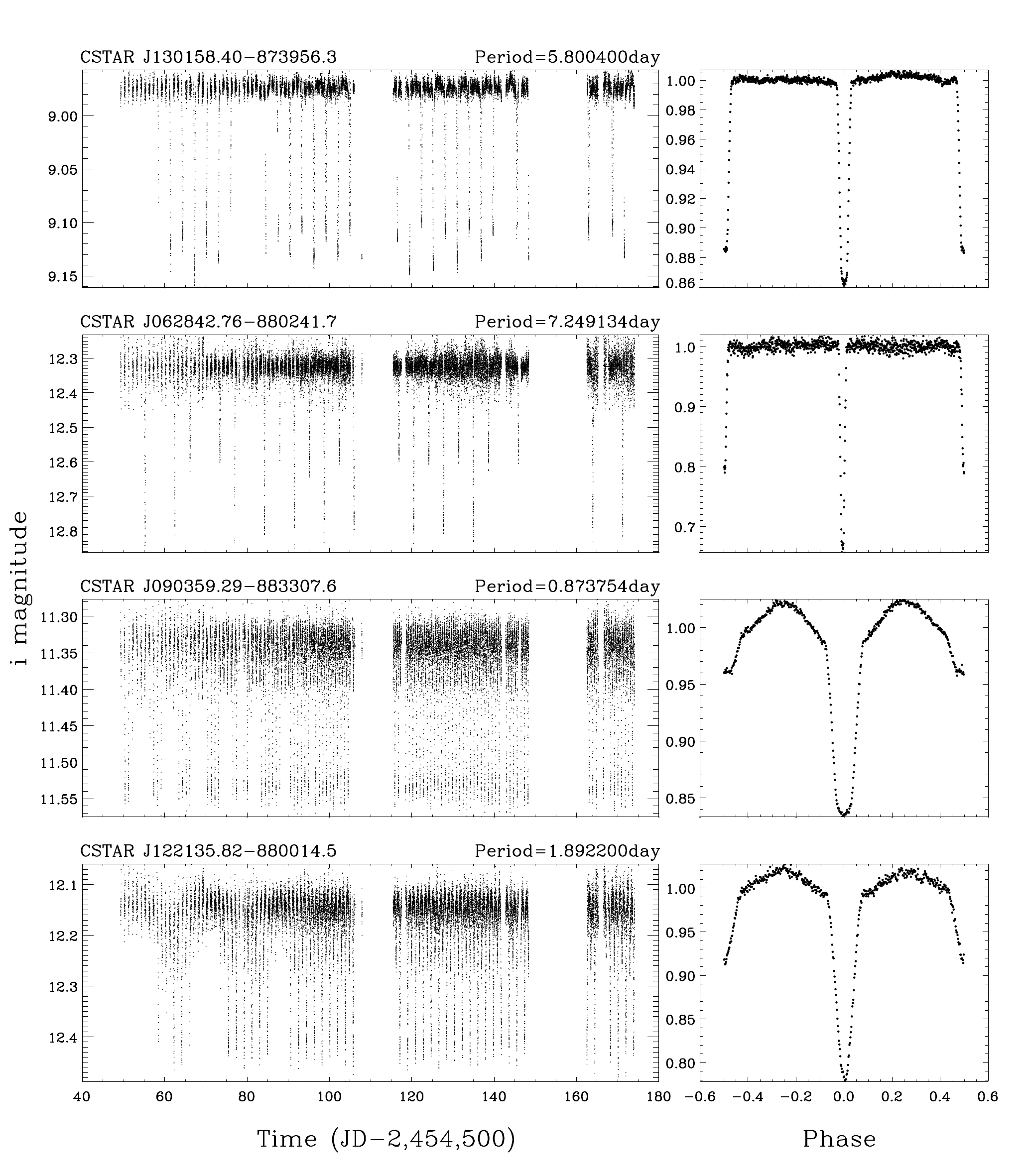}
\caption{  Example light curves of detached and semi-detach systems. The left-hand panels show the star brightness in magnitudes during the whole observation season. The right-hand panels show the phased and binned light curves in relative flux. CSTAR IDs and periods are given on top of every light curve.
\label{lc_ed_esd}}
\end{figure}

We fold the observed light curves to one period, normalize the flux and time, and fit the profiles with a bin size which is the same with the modelled light curves. Parameters are obtained rapidly by applying the trained EBAI matrix to the folded light curves, as shown in Tables \ref{ed_pars} and \ref{esd_pars}. The light curve of CSTAR J183057.87-884317.5 only shows primary eclipses. However, it has been found as an ED system with a high radial velocity amplitude of 12 km/s by \citet{wang14b}. We list it in the ED catalog but do not calculate its parameters. Figure \ref{lc_ed_esd} illustrates some ED and ESD light curves. The left-hand panels show the star brightness in magnitudes during the whole observation season. The right-hand panels show the phased and binned light curves in relative flux. Periods and CSTAR IDs are given on top of every left-hand panel.

\begin{deluxetable}{cccccccccc}
\tabletypesize{\tiny}
\tablewidth{0pt}
\tablecaption{Parameters of eclipsing detached binaries.}
\tablehead{
\colhead{CSTAR ID} & \colhead{mag}&\colhead{HJD}  & \colhead{Period} & \colhead{Crowding}& \colhead{$T_2/T_1$} & \colhead{$\rho_1+\rho_2$} & \colhead{$e\rm{sin}\omega$} & \colhead{$e\rm{cos}\omega$} & \colhead{$\sin i$} \\
 \colhead{} & \colhead{}& \colhead{(JD-2454500)}  & \colhead{(days)} & \colhead{} & \colhead{}& \colhead{}&  \colhead{}& \colhead{}& \colhead{}
}
\startdata
CSTAR J000116.84-874402.9	&	11.911	($\pm$	0.015	)&	51.595627	($\pm$	0.000812	)&	9.480642	($\pm$	0.000146	)&	0.045	&	0.994	&	0.076	&	0.051	&	0.126	&	1.000	\\
CSTAR J022528.30-875808.9	&	13.580	($\pm$	0.033	)&	55.026981	($\pm$	0.002729	)&	16.129131	($\pm$	0.001444	)&	0.016	&	0.091	&	0.085	&	-0.382	&	0.295	&	0.999	\\
CSTAR J031401.11-883253.1	&	14.491	($\pm$	0.070	)&	50.940662	($\pm$	0.001158	)&	1.208837	($\pm$	0.000021	)&	0.087	&	0.565	&	0.260	&	0.027	&	-0.003	&	1.000	\\
CSTAR J033130.30-880424.4	&	13.997	($\pm$	0.050	)&	47.732536	($\pm$	0.002857	)&	1.444413	($\pm$	0.000062	)&	0.006	&	0.752	&	0.234	&	-0.117	&	0.000	&	0.993	\\
CSTAR J061801.34-873953.9	&	13.371	($\pm$	0.031	)&	55.336460	($\pm$	0.017812	)&	6.048977	($\pm$	0.001605	)&	0.009	&	0.940	&	0.114	&	-0.040	&	0.003	&	0.998	\\
CSTAR J062842.76-880241.7	&	12.336	($\pm$	0.017	)&	55.236904	($\pm$	0.000568	)&	7.249134	($\pm$	0.000062	)&	0.095	&	0.596	&	0.117	&	0.056	&	0.011	&	1.000	\\
CSTAR J080846.28-880002.0	&	13.623	($\pm$	0.034	)&	50.397514	($\pm$	0.001443	)&	0.822784	($\pm$	0.000016	)&	0.006	&	0.574	&	0.423	&	0.005	&	0.004	&	0.945	\\
CSTAR J083940.85-873902.3	&	12.174	($\pm$	0.016	)&	57.158970	($\pm$	0.001962	)&	7.164405	($\pm$	0.000241	)&	0.011	&	0.357	&	0.079	&	-0.256	&	0.063	&	0.999	\\
CSTAR J090220.82-873741.0	&	12.355	($\pm$	0.017	)&	44.738434	($\pm$	0.004217	)&	1.627418	($\pm$	0.000089	)&	0.022	&	0.542	&	0.555	&	0.199	&	-0.023	&	0.943	\\
CSTAR J095836.02-882359.9	&	12.785	($\pm$	0.020	)&	50.778557	($\pm$	0.001152	)&	2.065943	($\pm$	0.000035	)&	0.025	&	0.461	&	0.289	&	0.006	&	0.003	&	0.976	\\
CSTAR J103232.73-882502.6	&	13.660	($\pm$	0.023	)&	53.738533	($\pm$	0.008547	)&	7.069140	($\pm$	0.000943	)&	0.015	&	1.045	&	0.109	&	0.138	&	-0.001	&	0.998	\\
CSTAR J104016.05-872929.8	&	11.102	($\pm$	0.013	)&	49.466064	($\pm$	0.000481	)&	0.868841	($\pm$	0.000006	)&	0.032	&	0.821	&	0.425	&	-0.029	&	-0.001	&	0.948	\\
CSTAR J130158.40-873956.3	&	8.984	($\pm$	0.004	)&	55.650978	($\pm$	0.000596	)&	5.800400	($\pm$	0.000059	)&	0.003	&	0.825	&	0.208	&	0.057	&	-0.002	&	0.998	\\
CSTAR J150537.60-873551.6	&	12.345	($\pm$	0.016	)&	52.568558	($\pm$	0.001329	)&	7.450763	($\pm$	0.000322	)&	0.088	&	0.647	&	0.125	&	0.112	&	0.001	&	0.997	\\
CSTAR J155705.55-873005.2	&	11.915	($\pm$	0.014	)&	51.797089	($\pm$	0.005051	)&	3.111666	($\pm$	0.000207	)&	0.004	&	0.500	&	0.230	&	-0.332	&	-0.011	&	0.993	\\
CSTAR J155917.54-880042.5	&	14.146	($\pm$	0.056	)&	48.417747	($\pm$	0.015527	)&	6.852954	($\pm$	0.001552	)&	0.013	&	0.629	&	0.106	&	0.086	&	0.002	&	0.999	\\
CSTAR J163136.82-874007.7	&	12.078	($\pm$	0.015	)&	58.014969	($\pm$	0.002532	)&	10.771624	($\pm$	0.000552	)&	0.006	&	0.988	&	0.061	&	0.054	&	-0.001	&	1.000	\\
CSTAR J183057.87-884317.5	&	9.839	($\pm$	0.007	)&	53.714790	($\pm$	0.002772	)&	9.922610	($\pm$	0.000403	)&	--	&	--	&	--	&	--	&	--	&	--	\\
CSTAR J193827.80-885055.9	&	12.884	($\pm$	0.021	)&	49.421947	($\pm$	0.001194	)&	6.752321	($\pm$	0.000111	)&	0.024	&	0.163	&	0.118	&	-0.139	&	-0.198	&	0.997	\\
CSTAR J200218.84-880250.0	&	11.977	($\pm$	0.015	)&	62.606724	($\pm$	0.003508	)&	19.141005	($\pm$	0.001177	)&	0.038	&	0.739	&	0.059	&	0.577	&	-0.357	&	1.000	\\
CSTAR J202830.07-874616.5	&	11.805	($\pm$	0.009	)&	51.365967	($\pm$	0.000346	)&	2.192940	($\pm$	0.000012	)&	0.000	&	0.417	&	0.340	&	0.196	&	-0.001	&	0.984	\\
CSTAR J205410.67-890348.2	&	10.030	($\pm$	0.008	)&	51.062473	($\pm$	0.001411	)&	1.857557	($\pm$	0.000038	)&	0.059	&	0.400	&	0.419	&	0.066	&	0.004	&	0.939	\\
CSTAR J224601.56-880459.2	&	13.823	($\pm$	0.040	)&	49.954193	($\pm$	0.003384	)&	7.760361	($\pm$	0.000409	)&	0.004	&	0.559	&	0.107	&	-0.013	&	0.005	&	0.999	\\
CSTAR J235727.17-882454.5	&	12.396	($\pm$	0.017	)&	50.875061	($\pm$	0.002693	)&	6.199004	($\pm$	0.000262	)&	0.008	&	0.291	&	0.131	&	0.007	&	-0.002	&	0.997	\\
	
\enddata
\label{ed_pars}
\begin{flushleft}
{\bf Note.} Columns 1-10 represent CSTAR ID, magnitude, the reference time of primary minimum, period, crowding, temperature ratio, the sum of fractional radii, the radial component of eccentricity, the tangential component of eccentricity, and the sine of inclination. J represents J2000.0. The errors of the physical parameters follow the distributions as shown in Fig \ref{lt_ed}. 
\end{flushleft}
\end{deluxetable}

\begin{deluxetable}{cccccccccc}
\tabletypesize{\tiny}
\tablewidth{0pt}
\tablecaption{Parameters of eclipsing semi-detached binaries.}
\tablehead{
\colhead{CSTAR ID} & \colhead{mag}&\colhead{HJD}  & \colhead{Period} & \colhead{Crowding} & \colhead{$T_2/T_1$} & \colhead{$\rho_1+\rho_2$} & \colhead{$e\rm{sin}\omega$} & \colhead{$e\rm{cos}\omega$} & \colhead{$\sin i$}\\
 \colhead{} & \colhead{}& \colhead{(JD-2454500)}  & \colhead{(days)} & \colhead{} & \colhead{}& \colhead{}&  \colhead{}& \colhead{}& \colhead{}
}
\startdata
CSTAR J074354.49-890737.3	&	12.538	($\pm$	0.018	)&	49.325817	($\pm$	0.000537	)&	0.797987	($\pm$	0.000006	)&	0.007	&	0.481	&	0.669	&	0.003	&	0.006	&	0.948	\\
CSTAR J084028.89-884700.4	&	13.807	($\pm$	0.039	)&	52.206177	($\pm$	0.029051	)&	13.027298	($\pm$	0.004186	)&	0.001	&	0.828	&	0.723	&	0.001	&	-0.003	&	0.913	\\
CSTAR J090359.29-883307.6	&	11.361	($\pm$	0.013	)&	49.491791	($\pm$	0.000177	)&	0.873754	($\pm$	0.000002	)&	0.000	&	0.596	&	0.677	&	0.000	&	0.000	&	0.871	\\
CSTAR J093334.26-865501.1	&	12.676	($\pm$	0.022	)&	52.293953	($\pm$	0.057799	)&	4.427421	($\pm$	0.003756	)&	0.022	&	0.748	&	0.609	&	0.000	&	0.000	&	0.987	\\
CSTAR J110803.52-870114.0	&	12.233	($\pm$	0.017	)&	49.716991	($\pm$	0.000752	)&	0.511559	($\pm$	0.000005	)&	0.018	&	0.506	&	0.615	&	0.066	&	0.021	&	0.915	\\
CSTAR J122135.82-880014.5	&	12.172	($\pm$	0.016	)&	50.905312	($\pm$	0.000439	)&	1.892200	($\pm$	0.000011	)&	0.002	&	0.582	&	0.613	&	0.000	&	0.000	&	0.913	\\
CSTAR J132349.26-881604.3	&	12.260	($\pm$	0.016	)&	51.587410	($\pm$	0.001441	)&	2.509739	($\pm$	0.000051	)&	0.003	&	0.835	&	0.689	&	0.000	&	0.000	&	0.907	\\
CSTAR J220502.55-895206.7	&	13.070	($\pm$	0.023	)&	79.778481	($\pm$	0.003821	)&	1.988110	($\pm$	0.000091	)&	0.000	&	0.473	&	0.575	&	0.039	&	0.000	&	0.919	\\
\enddata
\label{esd_pars}
\begin{flushleft}
{\bf Note.} Columns 1-10 represent CSTAR ID, magnitude, the reference time of primary minimum, period, crowding, temperature ratio, the sum of fractional radii, the radial component of eccentricity, the tangential component of eccentricity, and the sine of inclination. J represents J2000.0. The errors of the physical parameters follow the distributions as shown in Fig \ref{lt_ed}. 
\end{flushleft}
\end{deluxetable}

\subsection{ Parameters of Contact Binaries }
For ECs, there is no handle on eccentricity and argument of periastron. Instead, the lobe-filling configuration of a contact system links the Roche model with the radii of the components. As a result the photometric mass ratio $q$ can be estimated. In order to describe the contact degree, the fillout factor $F$ is given by \citet{prs11}:
\begin{equation}
F=\frac{\Omega - \Omega_{\rm L_2}}{\Omega_{\rm L_1} - \Omega_{\rm L_2}}
\end{equation}
where $\Omega$ \citep[see][Equation (1)]{wil71} is the surface potential of the common envelope, $\Omega_{\rm L_1}$ is the potential at the inner Lagrangian surface and $\Omega_{\rm L_2}$ is the potential at the outer Lagrangian surface. A star in a contact system will transfer mass to its companion through the the inner Lagrangian point ${\rm L_1}$, or lose mass through the the outer Lagrangian point ${\rm L_2}$. 

\begin{figure}
\epsscale{0.7}
\plotone{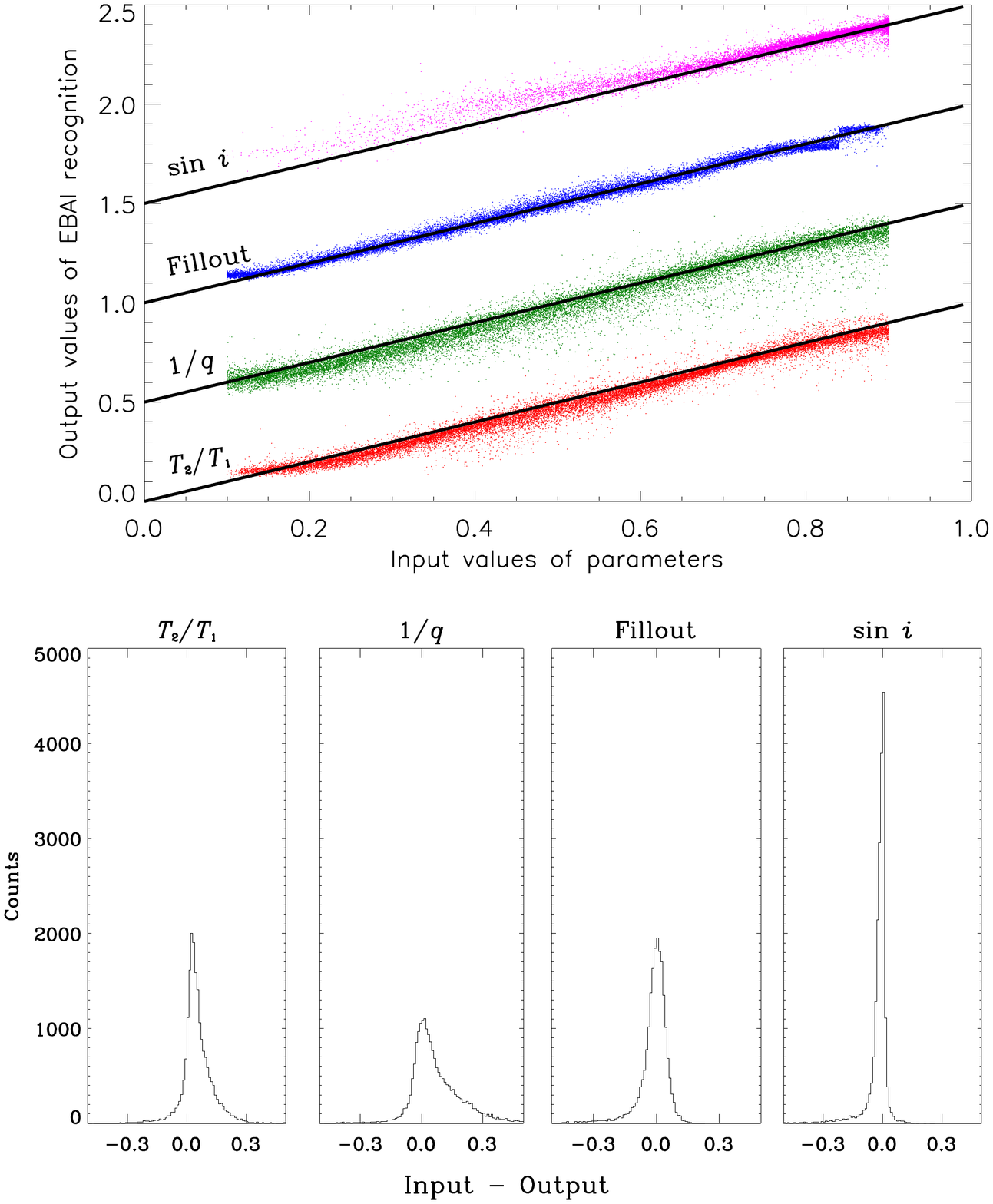}
\caption{ Similar as Figure \ref{lt_ed} but for ECs. The parameters are linearly scaled to the [0.1,0.9] interval by EBAI. Their actual ranges are given in Table \ref{bound}.
\label{lt_ec}}
\end{figure}

\begin{figure}
\epsscale{0.9}
\plotone{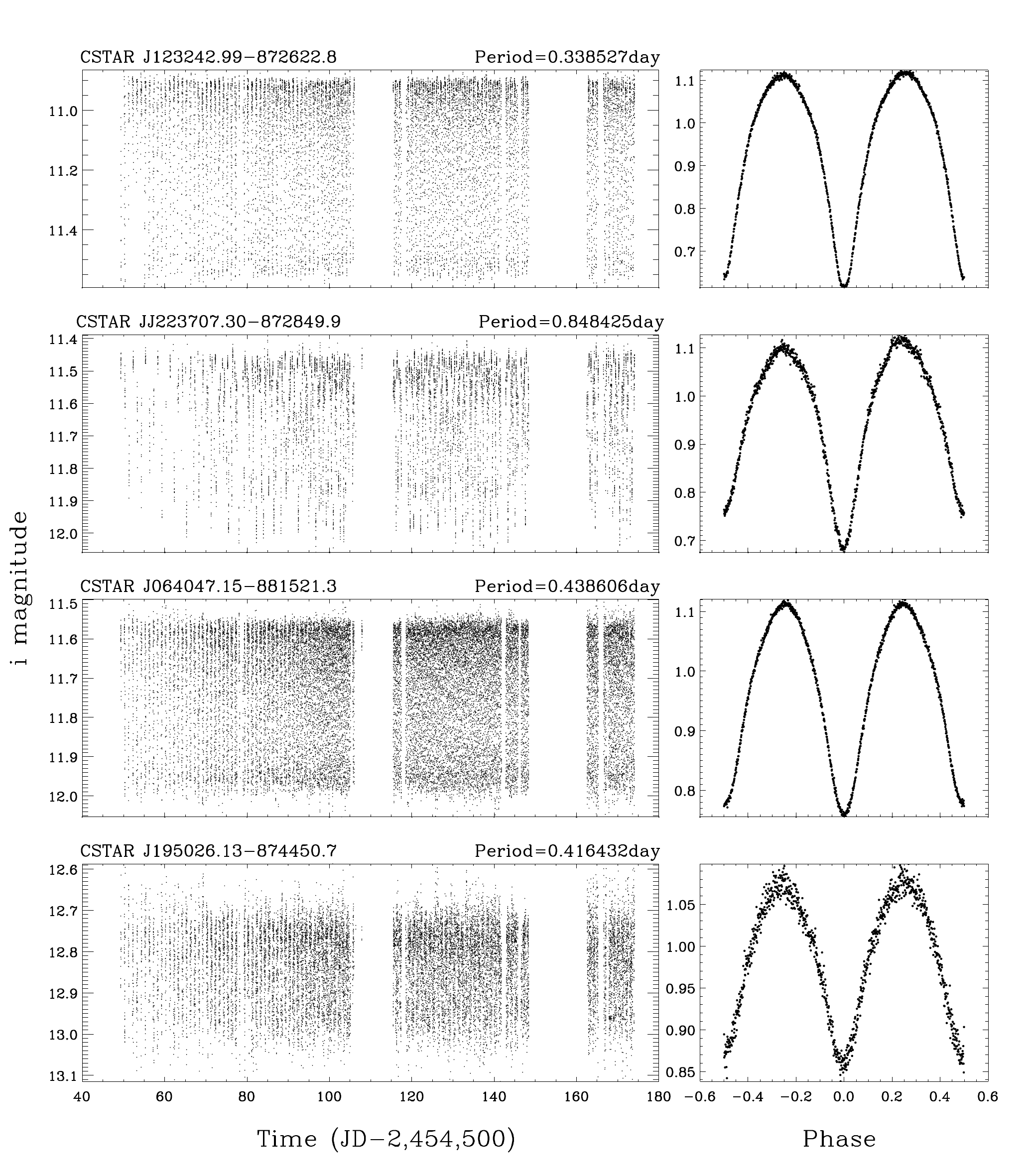}
\caption{ Example light curves of contact systems. CSTAR IDs and periods are given on top of every light curve.
\label{lc_ec}}
\end{figure}

From the above consideration, four principal parameters are chosen for EC: $T_{\rm{2}}\ /\ T_{\rm{1}}$, $q$, $F$, and ${\rm{\sin}}i$. The network is trained on $20,000$ simulated light curves for $1,000,000$ iterations. We test the correlation matrix with $10,000$ unknown EC light curves. The recognition results and the parameter error distributions are shown in Figure \ref{lt_ec}. We apply the correlation matrix to EC light curves of CSTAR. The parameters are given in Table \ref{ec_pars}. Figure \ref{lc_ec} shows some EC light curves. 

\begin{deluxetable}{ccccccccc}
\tabletypesize{\tiny}
\tablewidth{0pt}
\tablecaption{Parameters of eclipsing contact binaries.}
\tablehead{
\colhead{CSTAR ID} & \colhead{mag}&\colhead{HJD}  & \colhead{Period} & \colhead{Crowding} & \colhead{$T_2/T_1$} & \colhead{$M_2/M_1$} & \colhead{Fillout} &  \colhead{$\sin i$}\\
   \colhead{}& \colhead{}&\colhead{(JD-2454500)}  & \colhead{(days)} & \colhead{} &\colhead{}   &\colhead{} & \colhead{} & \colhead{} 
}
\startdata
CSTAR J005240.76-891732.4	&	13.997	($\pm$	0.044	)&	51.531921	($\pm$	0.000272	)&	0.292963	($\pm$	0.000002	)&	0.006	&	0.941	&	0.287	&	0.898	&	0.959	\\
CSTAR J031348.84-891511.7	&	13.047	($\pm$	0.023	)&	49.405579	($\pm$	0.000492	)&	0.344662	($\pm$	0.000002	)&	0.011	&	0.922	&	0.646	&	0.016	&	0.598	\\
CSTAR J042011.85-882503.5	&	12.647	($\pm$	0.019	)&	55.389160	($\pm$	0.000478	)&	0.395481	($\pm$	0.000003	)&	0.001	&	1.008	&	1.012	&	0.842	&	0.714	\\
CSTAR J051329.62-871942.6	&	11.939	($\pm$	0.016	)&	75.154015	($\pm$	0.000354	)&	0.384112	($\pm$	0.000003	)&	0.003	&	0.987	&	0.483	&	0.570	&	0.893	\\
CSTAR J051503.40-893226.6	&	14.431	($\pm$	0.079	)&	49.452221	($\pm$	0.001148	)&	0.358253	($\pm$	0.000006	)&	0.002	&	0.954	&	0.272	&	0.953	&	0.910	\\
CSTAR J061954.94-872047.5	&	12.399	($\pm$	0.019	)&	49.499104	($\pm$	0.001138	)&	0.491358	($\pm$	0.000008	)&	0.000	&	0.759	&	0.892	&	0.364	&	0.631	\\
CSTAR J064047.15-881521.3	&	11.721	($\pm$	0.014	)&	49.455452	($\pm$	0.000166	)&	0.438606	($\pm$	0.000001	)&	0.003	&	0.988	&	0.849	&	0.856	&	0.936	\\
CSTAR J071652.61-872856.4	&	13.472	($\pm$	0.035	)&	50.483288	($\pm$	0.000911	)&	0.383167	($\pm$	0.000005	)&	0.007	&	0.936	&	0.508	&	0.943	&	0.909	\\
CSTAR J073412.18-874037.3	&	13.218	($\pm$	0.021	)&	50.181190	($\pm$	0.000760	)&	0.331216	($\pm$	0.000003	)&	0.000	&	0.858	&	0.531	&	0.984	&	0.828	\\
CSTAR J084612.64-883342.9	&	11.997	($\pm$	0.015	)&	49.312592	($\pm$	0.000119	)&	0.267121	($\pm$	0.000000	)&	0.003	&	0.921	&	1.077	&	0.872	&	0.960	\\
CSTAR J123242.99-872622.8	&	11.100	($\pm$	0.013	)&	49.395302	($\pm$	0.000284	)&	0.338527	($\pm$	0.000001	)&	0.003	&	0.975	&	1.256	&	0.973	&	0.983	\\
CSTAR J124916.22-881117.6	&	13.834	($\pm$	0.040	)&	55.331085	($\pm$	0.000464	)&	0.352423	($\pm$	0.000002	)&	0.002	&	0.998	&	0.889	&	0.788	&	0.883	\\
CSTAR J135318.49-885414.6	&	12.783	($\pm$	0.020	)&	49.365437	($\pm$	0.000142	)&	0.266899	($\pm$	0.000001	)&	0.004	&	0.978	&	1.087	&	0.922	&	0.959	\\
CSTAR J142052.04-881433.4	&	12.599	($\pm$	0.018	)&	49.326588	($\pm$	0.000607	)&	0.400883	($\pm$	0.000003	)&	0.000	&	0.970	&	0.979	&	0.254	&	0.498	\\
CSTAR J142901.63-873816.2	&	13.542	($\pm$	0.032	)&	55.409655	($\pm$	0.000463	)&	0.348147	($\pm$	0.000003	)&	0.004	&	0.699	&	1.293	&	0.532	&	0.830	\\
CSTAR J181735.42-870602.2	&	10.819	($\pm$	0.012	)&	49.649178	($\pm$	0.002350	)&	0.352821	($\pm$	0.000012	)&	0.001	&	0.703	&	1.398	&	1.082	&	0.796	\\
CSTAR J195026.13-874450.7	&	12.832	($\pm$	0.021	)&	49.474827	($\pm$	0.000413	)&	0.416432	($\pm$	0.000002	)&	0.002	&	1.022	&	0.792	&	1.017	&	0.879	\\
CSTAR J223707.30-872849.9	&	11.624	($\pm$	0.014	)&	50.040791	($\pm$	0.001524	)&	0.848425	($\pm$	0.000017	)&	0.005	&	0.910	&	1.058	&	1.014	&	0.959	\\
\enddata
\label{ec_pars}
\begin{flushleft}
{\bf Note.} Columns 1-9 represent CSTAR ID, magnitude, the reference time of primary minimum, period, crowding, temperature ratio, mass ratio, fillout factor, and the sine of inclination. J represents J2000.0. The errors of the physical parameters follow the distributions as shown in Fig \ref{lt_ec}. 
\end{flushleft}
\end{deluxetable}

\section{ ECLIPSE TIMING VARIATIONS}

Companions are common around binary stars. It is suggested that close binaries are formed as a result of tidal friction and Kozai cycles in a multiple-star system \citep{bonnell01,fabrycky07}. Spectroscopic observations also support such scenario \citep{tokovinin97,tokovinin06}. The perturbation of the tertiary companion may change the eclipse mid-times. By calculating Eclipse Timing Variations (ETVs) we may find the tertiary companion. Other origins contributing to ETVs include star spots, mass transfer, spin-orbit transfer of angular momentum and orbit precession. Spin-orbit transfer of angular momentum is ignored in this paper because it changes the eclipse mid-times in the order of $10^{-5}$ of the orbital period \citep{app92}. Eccentricities of our short-period ECs and ESDs are close to zero due to tidal friction. Therefore their orbit precession can also be ignored. ETVs caused by star spots are discussed in Section 5.2. We do not calculate ETVs for EDs because of limited eclipses during the observation span.

\subsection{Computing Method of Eclipse Timing Variations}

To estimate the primary eclipse mid-times,  a simple linear increment is applied with known eclipse epoch and period. Then the whole light curve is divided into a lot of small segments. Each segment centres around a primary eclipse mid-time. We exclude segments with fewer points because such segments can reduce the fit precision. To get the eclipse template we fold all remaining segments. The template is fit using the following function \citep[see][equation (2)]{rappaport13} :
\begin{equation}
f = \alpha \left( t-t_0 \right)^2 + \beta \left( t-t_0 \right)^4 + f_0
\label{etv_fit_func}
\end{equation}
where f is the eclipse flux, $f_0$ is the minimum flux, t is the eclipse time and $t_0$ is the eclipse mid-time. Observed mid-times are obtained by comparing each segment with the theoretical template. All the other parameters except the eclipse mid-time are fixed the same as the template when comparing. A linear fit is applied to all the fitted mid-times. The differences between the fitted mid-times and their linear trend are ETVs. The same processes are applied to secondary eclipses to derive secondary ETVs. 

\subsection{Analysis of Eclipse Timing Variations}
The relationship between the primary and the secondary ETVs for a binary system may be correlated or anti-correlated. The correlation can be explained with a tertiary companion \citep{conroy13}; and the anti-correlation may be attribute to sun spots \citep{tran13}. However, when observed ETVs are less than one cycle, it is necessary to consider a probability of mass transfer. \citet{conroy13} adopt a Bayesian Information Criterion to distinguish the two cases.

To check the relationship between the primary and the secondary ETVs, we choose a similar method as searching planets \citep{ttvs3}. ETVs are fitted using the following function \citep[see][Equation(2)]{ming13}:
\begin{equation}
f = A {\rm{sin}} \left( \frac{2 \pi t}{P} \right) +  B {\rm{cos}} \left( \frac{2 \pi t}{P} \right) + C t + D \label{eq_f}
\end{equation}
where $A$, $B$, $C$, and $D$ are model parameters; and $P$ is the test period. $P$ is increased from 20 to 100 days with a step of 0.1 day. The degree of correlation is estimated by:
\begin{equation} \label{eq_xi}
\Xi(P) =  \frac{A_{p} A_{s}}{{\sigma}_{A_{p}} {\sigma}_{A_{s}}} + \frac{B_{p} B_{s}}{{\sigma}_{B_{p}} {\sigma}_{B_{s}}} 
\end{equation}
where the subscripts ``p" represents ``primary'' and ``s" represents ``secondary''; $\sigma_{A}$, $\sigma_{B}$, $\sigma_{C}$ and $\sigma_{D}$ are uncertainties of $A$, $B$, $C$, and $D$. The maximum $|\Xi|$ is adopted, which represents the degree of correlation. Positive $\Xi$ represents correlation and negative $\Xi$ represents anti-correlation. To confirm correlated ETVs, we calculate the false alarm probabilities (FAPs) by a bootstrap randomisation process: the ETVs are randomly scrambled $10^4$ times to obtain corresponding $\Xi_{\rm max}'$. The proportion of $\Xi_{\rm max}'$ larger than $\Xi_{\rm max}$ represents FAP. {\ym Induced periods are calculated to exclude the effect of sampling cadence. Finally two systems pass the FAP criteria of $10^{-3}$. CSTAR J084612.64-883342.9 has a FAP lower then $10^{-4}$ and CSTAR J220502.55-895206.7 has a FAP about $10^{-3.4}$. \citet{qia14} also claimed that a third body may exist aroud CSTAR J084612.64-883342.9.} 

\begin{figure}
\epsscale{0.6}
\plotone{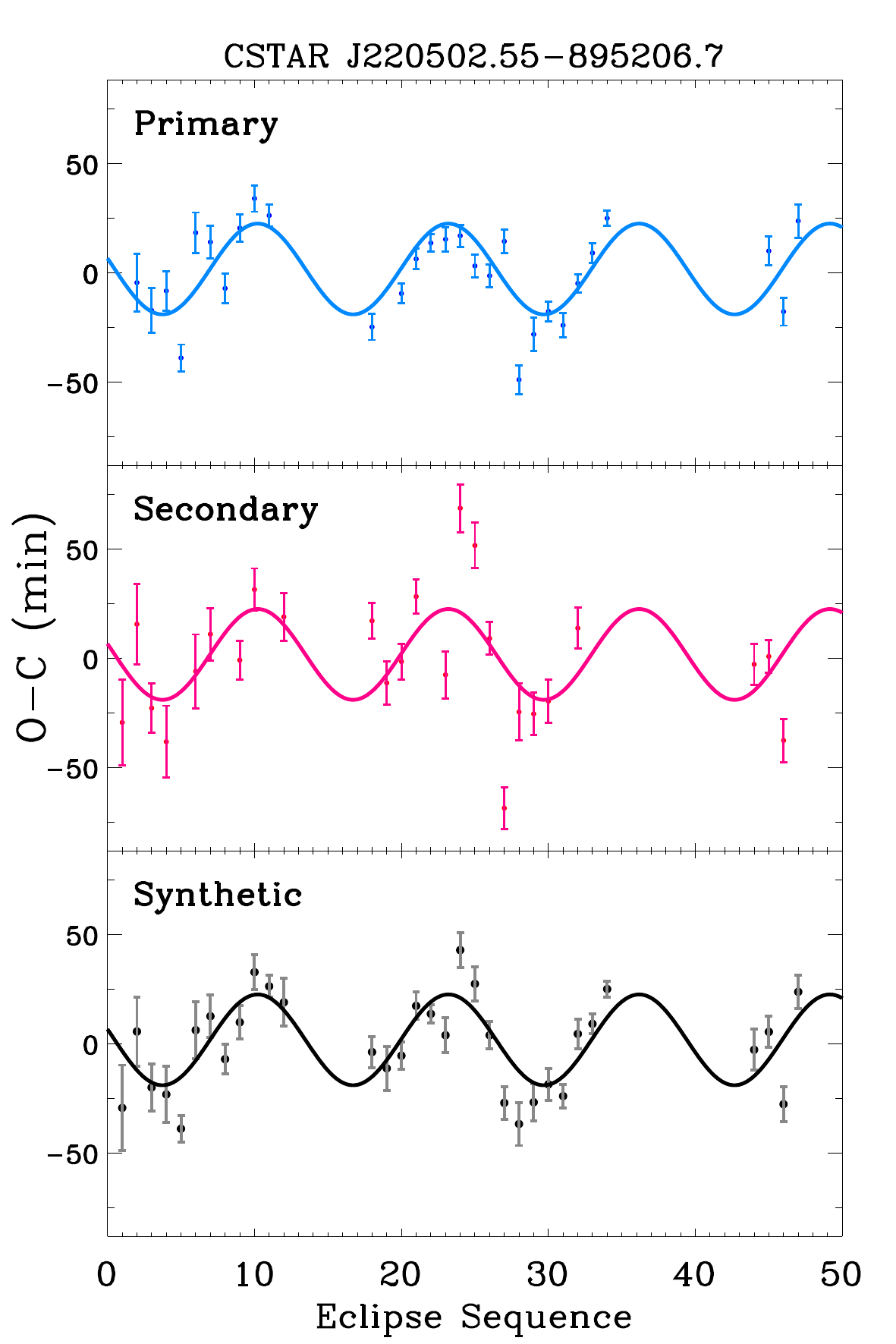}
\caption{ Primary (blue), secondary (red) and synthetic (black) ETVs of CSTAR J220502.55-895206.7. Synthetic ETV is the half of the sum of primary and secondary ETVs. The parameters of the third body are derived by fitting Equation (\ref{etv_equ}) with Synthetic ETV. ETV data are marked using filled circles with error bars, and the fit result of synthetic ETV is marked using thick line.  
\label{posetv}}
\end{figure}

\begin{figure}
\epsscale{0.9}
\plotone{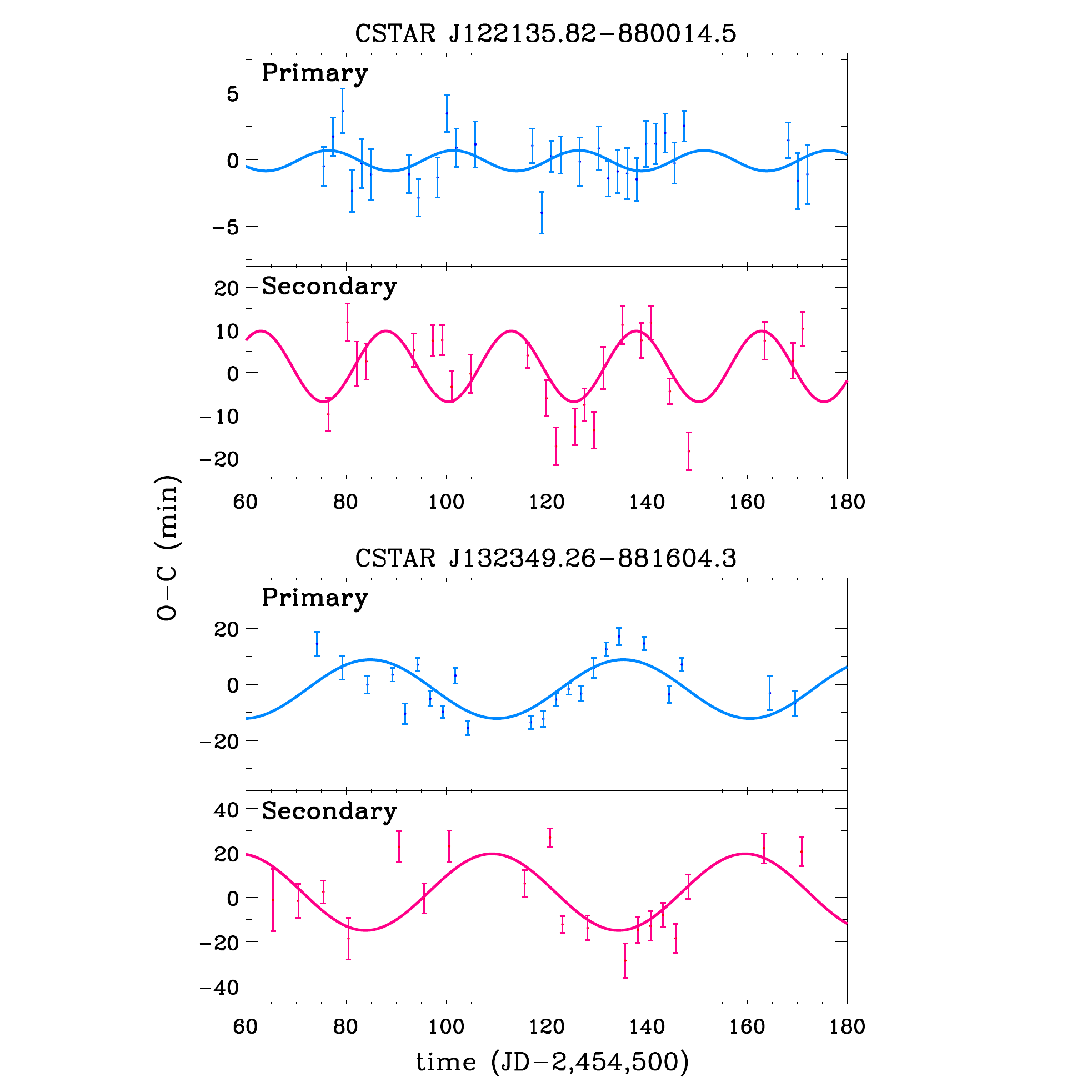}
\caption{Anti-correlated primary (blue) and secondary (red) ETVs of CSTAR J122135.82-880014.5 (top) and J132349.26-881604.3 (bottom). Primary and secondary ETVs are fitted using Equation (\ref{eq_f}), respectively. ETV data are marked using filled circles with error bars, and fit results are marked using thick lines.
\label{antietv}}
\end{figure}

Figure \ref{posetv} shows correlated ETVs of CSTAR J220502.55-895206.7 and Figure \ref{antietv} shows anti-correlated ETVs. For a correlated system, we sum its primary and secondary ETVs and divide the sum by two as the synthetic ETV of the binary system. {\ym If there is only primary eclipse or only secondary eclipse in one orbital period, then the ETV of the available eclipse is approximated as the synthetic ETV directly.} To derive the parameters of the third companion, the synthetic ETV is fitted using a triple-star model \citep{rappaport13,conroy13}:
\begin{equation}
	ETV = A \left[ \left( 1 - e_3^2 \right)^{1/2} \sin u_3(t) \cos \omega_3 + \left( \cos u_3(t) -e_3 \right) \sin \omega_3  \right]
\label{etv_equ}
\end{equation}
where
\begin{align}
	u_3(t) &= M_3(t) + e_3 \sin u_3(t) \\
	M_3(t) &= \left( t - t_0 \right) \frac{2 \pi}{P_3} \\
	A_{LTTE} &= \frac{G^{1/3}}{c (2 \pi )^{2/3}} \left[ \frac{m_3}{m_{123}^{2/3}} \sin i_3 \right] P_3^{2/3} 
\end{align}
where subscript ``3'' represents the third companion, $A$ is the amplitude, $e_3$ is eccentricity, $P_3$ is orbital period, $u_3(t)$ is eccentric anomaly, $M_3(t)$ is mean anomaly, $i_3$ is inclination, $\omega_3$ is argument of periastron, and $m_{123}$ is the mass of the whole system. The parameters we choose to give are $A$, $e_3$, $P_3$, and $\omega_3$. We fix the period corresponding to the maximum $\Xi$. Other parameters are changeable. For a more reliable fit, we choose Markov Chain Monte Carlo (MCMC) method instead of Levenberg-Marquardt algorithm. The results are shown in Table \ref{etv_pars}.

\begin{deluxetable}{lc}
\tabletypesize{\small}
\tablewidth{0pt}
\tablecaption{ Parameters of the potential companion around CSTAR J220502.55-895206.7.  }
\tablehead{
\colhead{Parameters} & \colhead{Values} 
}
\startdata
Period (days) & 25.36($\pm$0.83) \\
Amplitude (min) & 19.27($\pm$4.49)\\
eccentricity& 0.07($\pm$0.02) \\
w (rad)& 3.36($\pm$0.87) \\
FAP (log)& -3.4 
\enddata
\label{etv_pars}
\end{deluxetable}

\section{RESULTS}
We identify and classify $53$ eclipsing binaries, containing $24$ EDs, $8$ ESDs, $18$ ECs, and $3$ ELLs. The distributions of their physical parameters are shown in Figure \ref{hist_ec_ed}. Some EDs may be in the dynamically hot stage because of high eccentricities. We plot four EDs with eccentricities higher than 0.1 (CSTAR J$000116.84$-$874402.9$, J$200218.84$-$880250.0$, J$083940.85$-$873902.3$, and J$193827.80$-$885055.9$) in Figure \ref{large_e} and give the period-eccentricity diagram in Figure \ref{ed_p_e}. Many folded light curves appear asymmetry in brightness as shown in Figure \ref{large_asym}. ETV analyses present systems with correlation and anti-correlation between the primary and the secondary eclipses, respectively. The systems mentioned above are discussed as follows. 

\subsection{Binary Parameters and Typical Characteristics}
\begin{figure}
\epsscale{0.7}
\plotone{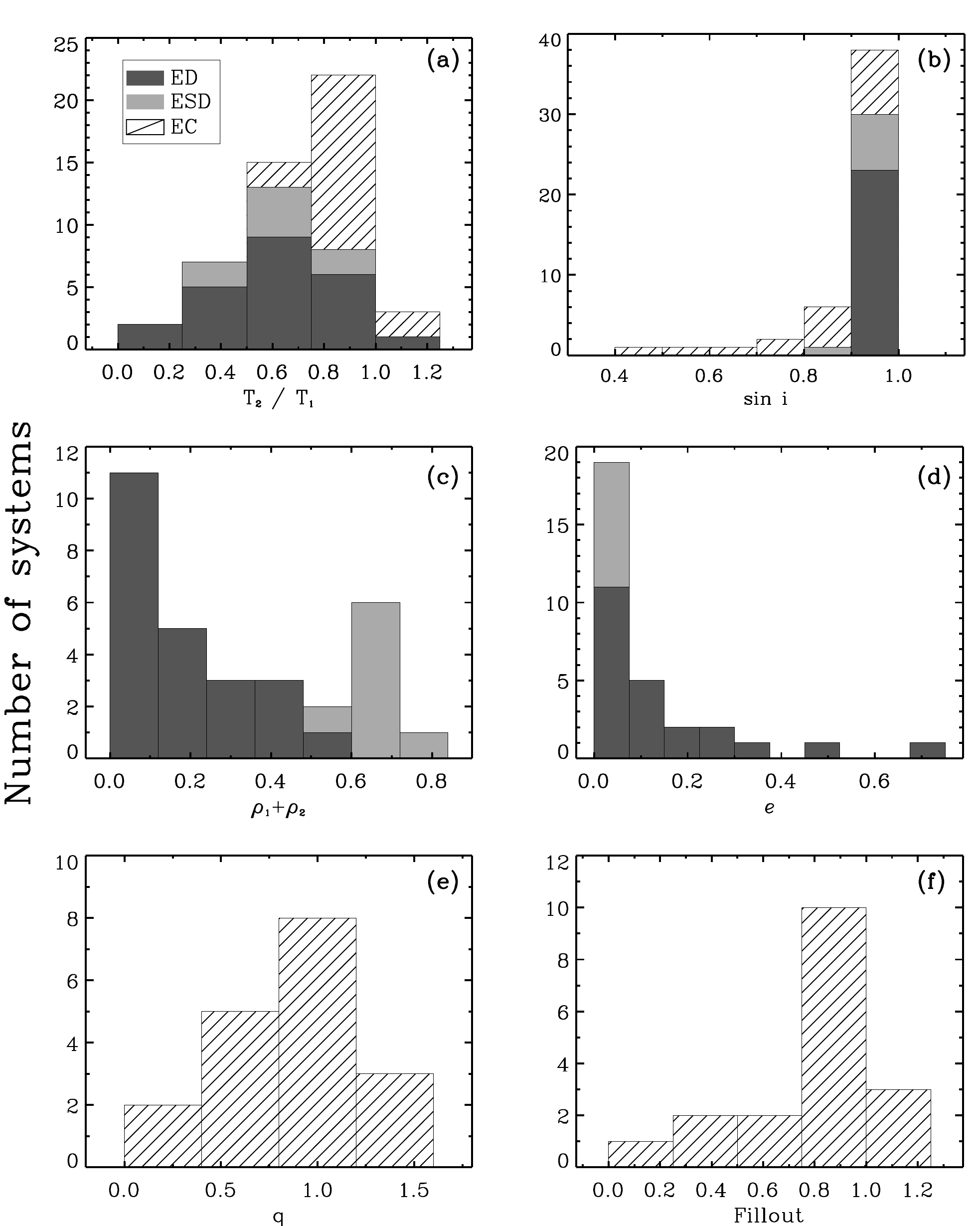}
\caption{Statistics of the number of eclipsing binaries with different physical parameters: (a) temperature ratio; (b) the sine of orbital inclination; (c) the sum of fractional radii; (d) eccentricity; (e) mass ratio, and (f) fillout factor.
\label{hist_ec_ed}}
\end{figure}

\begin{figure}
\epsscale{0.8}
\plotone{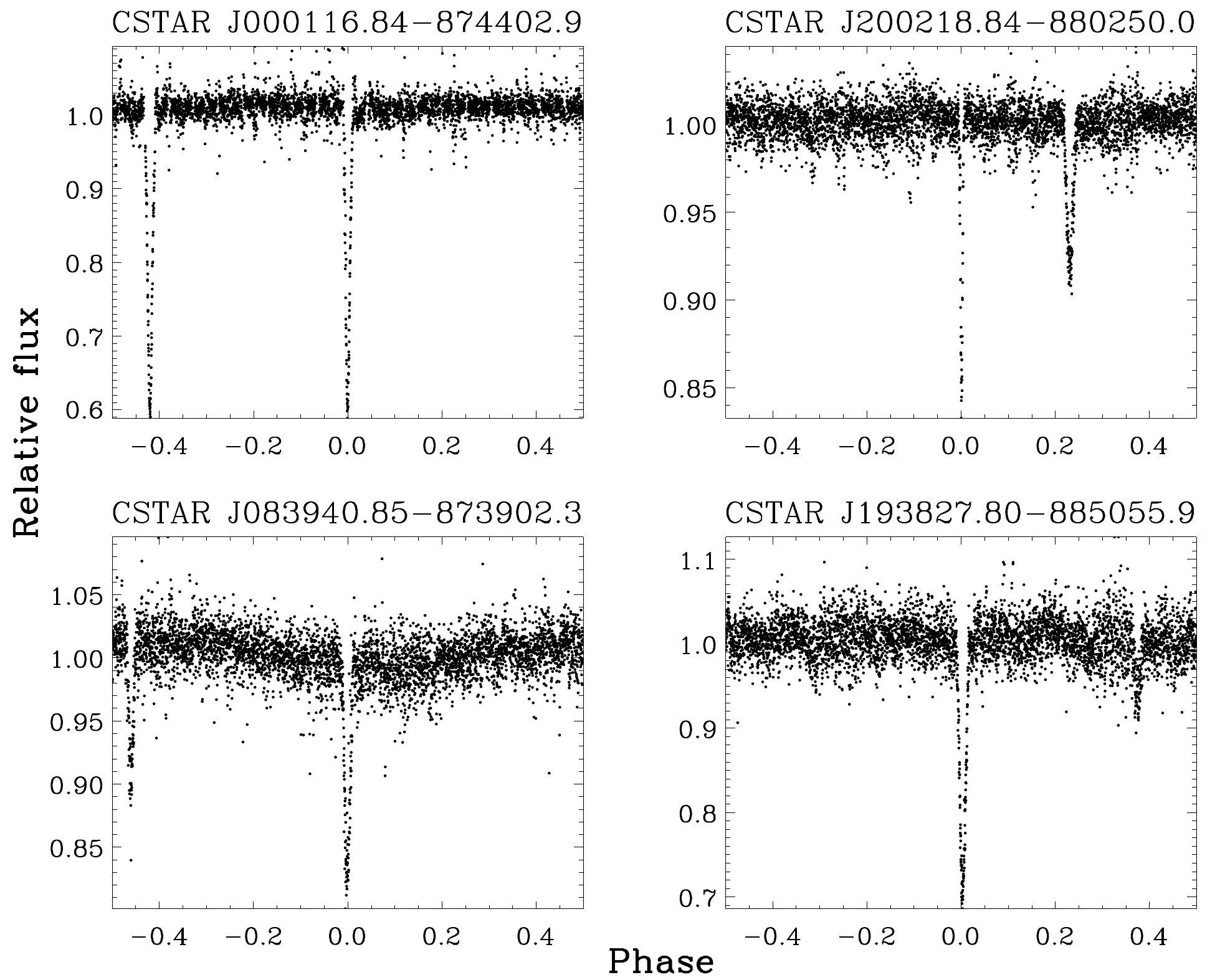}
\caption{ Example light curves of eccentric detached binaries. The light curves are folded and binned to 5,000 points. CSTAR IDs are given on top of each panel.
\label{large_e}}
\end{figure}

Eclipsing binary parameters are given in Tables \ref{ed_pars}-\ref{ec_pars}. The distributions of the physical parameters (see Table \ref{bound}) are shown in Figure \ref{hist_ec_ed}. Because samples in this paper are limited, we use a large bin size to test some typical characteristics. 

For detached and semi-detached systems, the eclipse depth ratio reflects the temperature ratio. In the condition of equal depth eclipses its value is 1. The temperature ratios of the EDs and ESDs in this paper center around the value lower than 1. \citet{prs11} explain that it is because the orbital eccentricity and star radii can affect the eclipse depth, thus increasing the scatter when the temperature ratio approaches to 1. The inclinations are close to $90^\circ$ for detached and semi-detached systems. This is because eclipses can only be seen in the edge-on geometrical configuration when the two components are not very close. The sum of the fractional radii distribute around $0.1$ for EDs and $0.6$ for ESDs. 

\begin{figure}
\epsscale{0.8}
\plotone{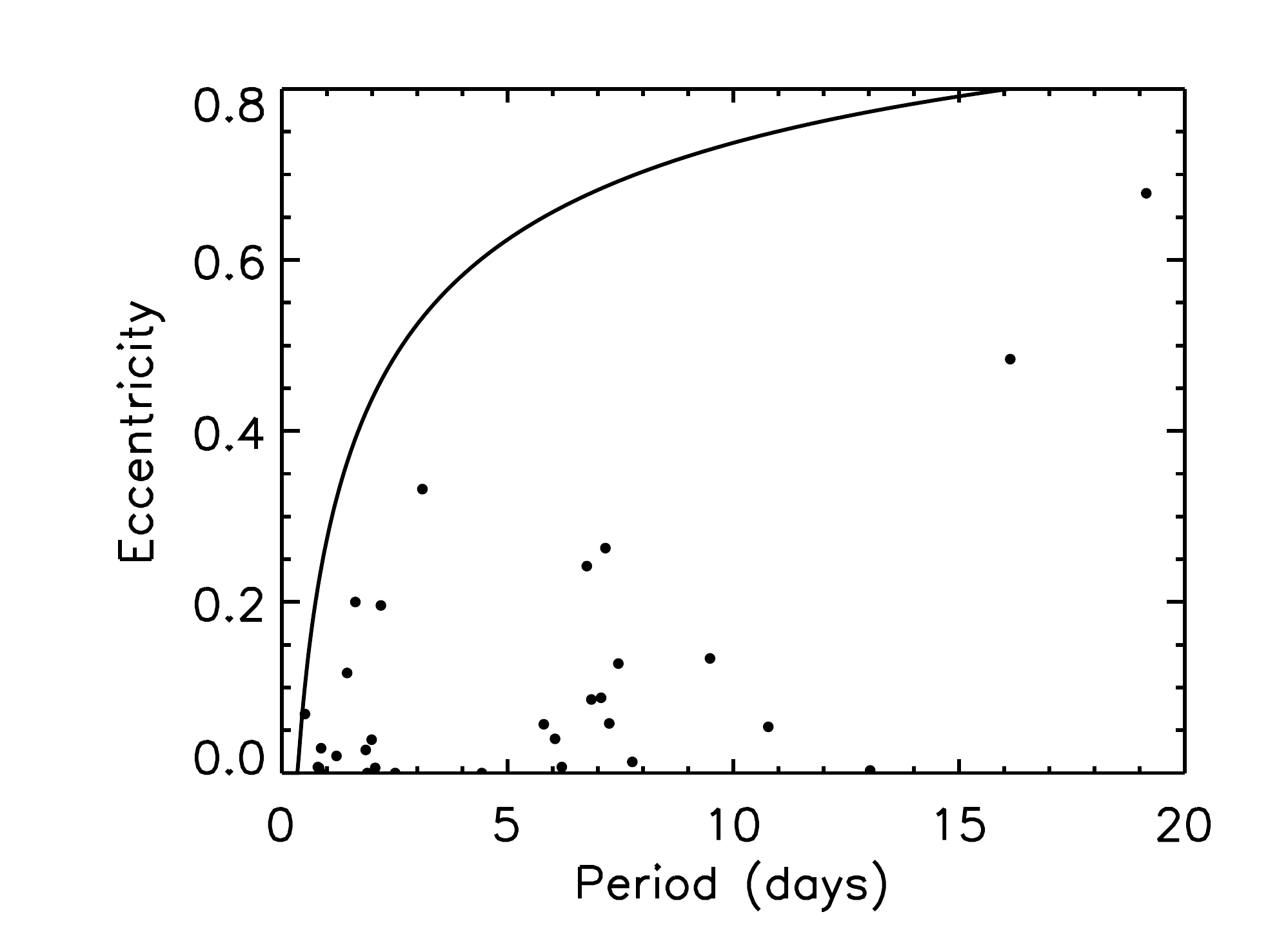}
\caption{ The eccentricities and the orbital periods of the detached and semi-detached binaries. The black line is the upper envelope derived by \citet{maz08} to constrain the binary region. All the 2751 binaries from the SB9 catalog are used to obtain the expression of the upper envelope. EDs and ESDs in this paper are marked with black dots.
\label{ed_p_e}}
\end{figure}

What is interesting for EDs and ESDs is their eccentricity distribution. Most of the eccentricities are close to zero. However, the highest eccentricity can reach 0.679 (CSTAR J$200218.84$-$880250.0$). \citet{hut81} analyzes the tidal evolution of binaries and concludes that the time scale of circularization can be a relatively slow process. \citet{maz08} plots the eccentricities as a function of the orbital periods using all the 2751 binaries from the official IAU catalog of spectroscopic binaries (SB9). He derives an ``upper envelope'' to constrain the binary eccentricity \citep[see][Equation (4.4)]{maz08}:
\begin{equation}
f(P) = E - A \exp(-(pB)^C)
\end{equation}
where $E=0.98, A=3.25, B=6.3$, and $C=0.23$. The EDs and ESDs in this paper are all below the upper envelope as shown in Figure \ref{ed_p_e}. Two detached systems (CSTAR J$200218.84$-$880250.0$ and J$022528.30$-$875808.9$) at the upper right of Figure \ref{ed_p_e} have longer periods and larger eccentricities. Such systems have experienced less of the circularization process. Therefore they are important to investigate how circularization process can affect the binary components by comparing high-eccentricity binaries with circular binaries \citep{shi14}.

For contact systems, the two lobe-filling components can transfer mass to each other through the inner Lagrangian point. They are easy to reach a thermal contact because of the common envelope. Therefore their temperature ratios center around 1 as shown in Figure \ref{hist_ec_ed}(a). The relatively larger size of the Roche lobe also relax the limitation of edge-on geometrical requirement. ECs with lower inclinations can be detected as shown in Figure \ref{hist_ec_ed}(b). The photometric mass ratios of ECs peak at 1. The fillout factors for the ECs are not roughly uniform because overcontact systems and close-to-contact systems contribute to the peak at 1.

\subsection{Light Curves with O'Connell Effect}

\begin{figure}
\epsscale{0.8}
\plotone{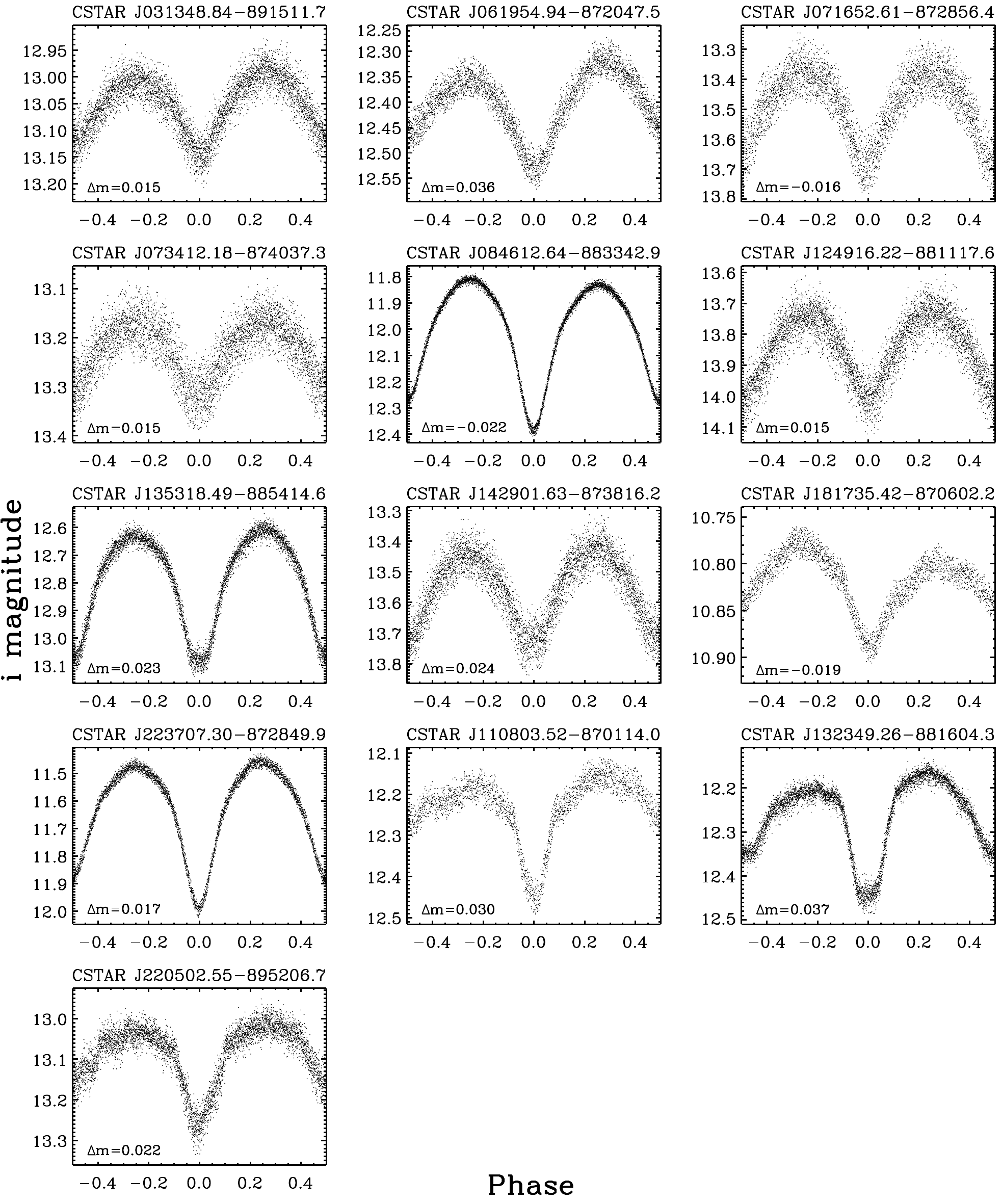}
\caption{  Light curves with O'Connell effect $|\Delta m| \geq  0.01$. $\Delta m=m_{\rm II} - m_{\rm I}$ in the lower left of each panel represents the difference between the maxima, where $m_{\rm I}$ is the peak magnitude after primary minimum and $m_{\rm II}$ is the peak magnitude after secondary minimum. The light curves are folded and binned to 5,000 points. CSTAR IDs are given on top of each panel.
\label{large_asym}}
\end{figure}

Many CSTAR phased light curves of ECs and ESDs appear different maxima in brightness as shown in Figure \ref{large_asym}. These light curves have been phased and binned to 5000 points equally spaced. Such phenomenon is called O'Connell effect \citep{oco51,dav84}. The O'Connell Effect can be quantitatively expressed by measuring the difference between the two out-of-eclipse maxima
\begin{equation}
\Delta m = m_{\rm II} - m_{\rm I}
\end{equation}
where $m_{\rm I}$ and $m_{\rm II}$ are the peak magnitude after primary minimum and secondary minimum, respectively. To derive $m_{\rm I}$ and $m_{\rm II}$, we fold each light curve with its orbital period and fit the parts centered around each maximum using Equation (\ref{etv_fit_func}).

Table \ref{oconnell} presents the contact and semi-detached systems with $|\Delta m|$ greater than 0.01. Nearly half of the contact systems are listed in this table. Therefore O'Connell Effect is common among eclipsing binary systems. The reasons for O'Connell effect are still debatable. \citet{mul75} proposed that large toroidal magnetic fields may be generated because of enforced rapid rotation in deep convection zones of contact binaries. The toroidal magnetic fields can cause active low temperature zones, namely spots. The spots can be used to explain the reason of O'Connell effect. However, not all the ESDs have deep convection zones like the ECs. Instead, they may have hot spots because of mass transfer. In addition, O'Connell effect can also be found in some EDs. \citet{dav84} have a discovery that the O'Connell effect of detached systems is correlated with the color index significantly. Therefore O'Connell effect may be caused by more than one mechanism. More eclipsing binaries with different color index are needed to study it.
 
\begin{deluxetable}{cccccc}
\tabletypesize{\scriptsize}
\tablewidth{0pt}
\tablecaption{Contact and semi-detached systems with O'Connell effect}
\tablehead{
\colhead{CSTAR ID} & \colhead{$\Delta m$}&\colhead{Type}  & \colhead{CSTAR ID} & \colhead{$\Delta m$} & \colhead{Type} \\
   \colhead{}& \colhead{(mag)}&\colhead{}  & \colhead{} & \colhead{(mag)} &\colhead{} 
}
\startdata
CSTAR J031348.84-891511.7	&	0.015	&	EC	&		CSTAR J061954.94-872047.5	&	0.036	&	EC	\\
CSTAR J071652.61-872856.4	&	-0.016	&	EC	&		CSTAR J073412.18-874037.3	&	0.015	&	EC	\\
CSTAR J084612.64-883342.9	&	-0.022	&	EC	&		CSTAR J124916.22-881117.6	&	0.015	&	EC	\\
CSTAR J135318.49-885414.6	&	0.023	&	EC	&		CSTAR J142901.63-873816.2	&	0.024	&	EC	\\
CSTAR J181735.42-870602.2	&	-0.019	&	EC	&		CSTAR J223707.30-872849.9	&	0.017	&	EC	\\
CSTAR J110803.52-870114.0	&	0.030	&  ESD  &		CSTAR J132349.26-881604.3	&	0.037	&	ESD	\\  
CSTAR J220502.55-895206.7	&	0.022	&  ESD  &			&		&		\\
\enddata
\label{oconnell}
\begin{flushleft}
{\bf Note.} $\Delta m=m_{\rm II} - m_{\rm I}$, where $m_{\rm I}$ is the peak magnitude after primary minimum and $m_{\rm II}$ is the peak magnitude after secondary minimum.
\end{flushleft}
\end{deluxetable}

\subsection{Eclipse Timing Variations for Binaries}

ETVs are useful for studying the apsidal motion, mass transfer and loss, solar-like activities in late-type stars, light-time effect of a third companion, etc. They are helpful in understanding the formation and evolution of binary systems. Light curves of short-period binaries from the Antarctica are available for tens of days thanks to the polar nights. Therefore their ETVs are more continuous.

We only calculate ETVs for semi-detached and contact binaries. Spin-orbit 
transfer of angular momentum \citep{app92} does not create observable 
signals because their orbital periods are very short and the best 
precision of our ETVs is 1 min as shown in Figure \ref{posetv}.  Apsidal motion is 
ignored because the eccentricities of semi-detached and contact binaries 
are close to zero.

Systems with potential large spots are shown in Figure \ref{antietv}. For each system, we can see obvious anti-correlation between the primary and the secondary eclipses. {\ym We find two eclipsing binary systems with a potential third body. They are CSTAR J084612.64-883342.9 and CSTAR J220502.55-895206.7. Recently \citet{qia14} also claimed that CSTAR J084612.64-883342.9 is a triple system  and derived the parameters of the third body with more observations \citep[see][Table 5]{qia14}. In this paper we analysis the other system CSTAR J220502.55-895206.7. The ETVs of CSTAR J220502.55-895206.7} are given in Figure \ref{posetv}. Parameters of the close-in third body are given in Table \ref{etv_pars}. {\ym The orbital period of the third body is 25.36 day, indicating a close distance from the binary.} Because the mass of the third companion and the orbital inclination are coupled in the amplitude, we cannot tell the nature of the third companion.

\section{CONCLUSIONS}

CSTAR, with an aperture of $14.5$ cm (effective aperture 10 cm), was fixed to point in the direction near the Celestial South Pole at Dome A. After analyzing $i$-band data of CSTAR observed in 2008, a master catalog containing $22,000$ sources was obtained. There are about $20,000$ sources between $8.5$ mag and $15$ mag. The polar night condition of Dome A and the large aperture of CSTAR make it more suitable to search and analyze eclipsing binaries.  

In this work, we analyze each light curve using the Lomb-Scargle, Phase Dispersion Minimization, and Box Least Squares method to search variables. To pick out binaries from the variables, the period, the CCD position, and the morphology of the light curves are compared and checked. Finally, we discover ${\ym 53}$ eclipsing binaries in the field of view of CSTAR. Therefore, the binary occurrence rate is ${\ym 0.26\%}$ concerning the $\sim20,000$ sources in the master catalog. It is lower compared with $0.8\%$ of Hipparcos and $1.2\%$ of  {\it Kepler}, but close to OGLE binary occurrence rate of $0.2\%$ for all stars. There are more EDs and ESDs than ECs, as indicated by  {\it Kepler}. 

The parameters of the eclipsing binaries are calculated by using the PHOEBE package and EBAI pipeline. For different types of eclipsing binaries, we choose different parameters. The general statistical characteristics of the parameters are similar with {\it KEPLER}. Since the number of the eclipsing binaries in this paper is limited, it is not possible to investigate detailed statistical characteristics. However, individual systems on the edge of the parameter distributions are still of concern. Some detached binaries are found to be very eccentric. We check their eccentricities in the period-eccentricity diagram offered by \citet{maz08}, and the eccentricities are all within the restricted area. Eccentric binary orbit indicates dynamical hot stage. Therefore such systems are valuable to study the evolution of binaries and the impact of circularization process on binary components.

Antarctic polar nights also offer good opportunities to investigate light curves detailedly and continuously. For short-period variables, observations can be taken during the whole orbital period without interruption. Therefore, it is very efficient to analyze the variations of some physical quantities, such as ETV. We calculate the ETVs for all semi-detached and contact systems. The precision of CSTAR ETVs can achieve one minute at $9$ mag, which can reveal the existence of a massive third companion. The ETV analyses present (1) two systems with correlated primary and secondary ETVs, implying potential companions; and (2) another two systems with anti-correlated primary and secondary ETVs, implying star spots. The orbital parameters of the third boy in system CSTAR J220502.55-895206.7 are derived using a triple-star dynamical model.

\acknowledgments
We are grateful to Andrej Pr{\v s}a for helpful discussions and for sharing his PHOEBE script with us. We thank Xiaobin Zhang and Changqing Luo for their helpful comments. {\ym We sincerely appreciate the crew involved in the CSTAR project and the Chinese Antarctic Science team who delivered and set up CSTAR at Dome A for the first time. We are also grateful to the High Performance Computing Center (HPCC) of Nanjing University for doing the numerical calculations in this paper on its IBM Blade cluster system.} This research has been supported by the Key Development Program of Basic Research of China (No. 2013CB834900), the National Natural Science Foundations of China (Nos. 10925313, 11003010 and 11333002),  Strategic Priority Research Program ``The Emergence of Cosmological Structures" of the Chinese Academy of Sciences (Grant No. XDB09000000), the Natural Science Foundation for the Youth of Jiangsu Province (No. BK20130547), Jiangsu Province Innovation for PhD candidate (No. KYZZ\_0030 and KYLX\_0031), 985 Project of Ministration of Education and Superiority Discipline Construction Project of Jiangsu Province.

\end{CJK*}

\end{document}